\documentclass[a4paper,onecolumn,11pt,accepted=2023-09-04]{quantumarticle}
\pdfoutput=1
\usepackage{amsmath}
\usepackage{amsfonts}
\usepackage{amsthm}
\usepackage{amssymb}
\usepackage{graphicx}
\usepackage{enumerate}
\usepackage{color}
\usepackage[T1]{fontenc}
\usepackage{braket}
\usepackage{todonotes}
\usepackage{soul}
\usepackage{subfigure}
\usepackage{mathrsfs}
\usepackage{float}
\usepackage{bm}
\usepackage{multirow}

\graphicspath{{figures/}}

\DeclareMathOperator{\Tr}{Tr}

\newtheorem{proposition}{Proposition}

\newtheorem*{theorem*}{Theorem}
\newtheorem*{proposition*}{Proposition}

\usepackage[bookmarks=false,colorlinks,citecolor=blue,
linkcolor=blue,anchorcolor=blue,urlcolor=blue
]{hyperref}
\begin{document}

\title{Tilted Hardy paradoxes for device-independent randomness extraction}

\author{Shuai Zhao}
\affiliation{Department of Computer Science, The University of Hong Kong, Pokfulam Road, Hong Kong}
\orcid{0000-0002-8720-5476}
\author{Ravishankar Ramanathan}
\email{ravi@cs.hku.hk}
\affiliation{Department of Computer Science, The University of Hong Kong, Pokfulam Road, Hong Kong}
\orcid{0000-0003-1119-8721}
\author{Yuan Liu}
\email{yuan59@connect.hku.hk}
\affiliation{Department of Computer Science, The University of Hong Kong, Pokfulam Road, Hong Kong}
\author{Pawe{\l} Horodecki}
\affiliation{International Centre for Theory of Quantum Technologies, University of Gda\'{n}sk, Wita Stwosza 63, 80-308 Gda\'{n}sk, Poland}
\affiliation{Faculty of Applied Physics and Mathematics, Gda\'{n}sk University of Technology, Gabriela Narutowicza 11/12, 80-233 Gda\'{n}sk, Poland}
\orcid{0000-0002-3233-1336}

\begin{abstract}
The device-independent paradigm has had spectacular successes in randomness generation, key distribution and self-testing, however most of these results have been obtained under the assumption that parties hold trusted and private random seeds. In efforts to relax the assumption of measurement independence, Hardy's non-locality tests have been proposed as ideal candidates. In this paper, we introduce a family of tilted Hardy paradoxes that allow to self-test general pure two-qubit entangled states, \textcolor{black}{as well as certify up to $1$ bit of local randomness.
We then use these tilted Hardy tests to obtain an improvement in the generation rate in the state-of-the-art randomness amplification protocols for Santha-Vazirani (SV) sources with arbitrarily limited measurement independence.
Our result shows that device-independent randomness amplification is possible for arbitrarily biased SV sources and from almost separable states.}
Finally, we introduce a family of Hardy tests for maximally entangled states of local dimension $4, 8$ \textcolor{black}{as the potential candidates for DI randomness extraction to certify up to the maximum possible $2 \log d$ bits of global randomness.}
\end{abstract}
\maketitle

\section{Introduction}
One of the most fundamental features of quantum mechanics is the presence of correlations that cannot be explained by any local hidden variable theory \cite{einstein1935can,schrodinger1935discussion}. Apart from being of fundamental interest, this phenomenon of Bell non-locality has led to the powerful idea of device-independent (DI) quantum key distribution \cite{barrett2005no,acin2007device}, randomness generation \cite{pironio2010random,pironio2013security} and certification of quantum systems \cite{mayers1998quantum,mayers2004self}. 

The strength of the device-independent paradigm is that no assumption on the nature of the systems subject to measurement needs to be made. Indeed, one may simply consider the systems participating in the Bell experiment to be (two or more) black boxes that parties provide an input to and obtain an outcome from, not taking into account the complex details of the physical implementation at all. The observation of a Bell inequality violation then allows one to make nontrivial deductions about the nature of the systems under study, such as the presence of entanglement, or a lower bound on the system dimension, or the non-determinism of the measurement outputs. In the extreme case, observation of maximum violation of certain Bell inequalities even permits the device-independent certification (\textit{self-testing}) \cite{mayers2004self,Supic2020selftestingof} of the quantum state and measurements performed on the system, i.e., their uniqueness up to irrelevant local equivalences. 

Such self-testing has obvious advantages over traditional certification methods such as those based on quantum tomography \cite{goh2019experimental}, and a lot of attention has therefore been devoted recently to designing Bell inequalities suited for self-testing different entangled quantum states. However, DI certification based on the violation of Bell inequalities nevertheless still relies upon some assumptions, the foremost being the requirements of no-signaling between the local systems (typically enforced by space-like separation), and that the measurements on the local systems are chosen freely and randomly. 

This latter requirement known as measurement independence is typically justified by the assumption that the parties hold independent and trusted random number generators (i.e., trusted and private random seeds). However, this assumption is clearly incongruous with the very framework of device-independence, wherein all devices held by the honest parties may have been tampered with, or even provided by, an adversary. To elaborate, consider the adversarial scenario wherein an adversary Eve has had access to the very devices used by the honest parties in the protocol. If such an adversary was able to influence the local random number generators held by the parties, then she would be able to ensure that the parties only hold imperfect seeds (about which Eve has some side information). In the extreme case, Eve may even be able to prepare devices that only operate according to local hidden variable behaviors, and yet appear non-local to the parties due to the imperfection of their seeds. Indeed, when no measurement independence at all is available, one cannot demonstrate any non-locality. It is therefore of vital importance to extend the studies on device-independent certification (as well as other tasks such as key distribution and random number generation) to the scenario in which only limited (arbitrarily small) measurement independence is available \cite{colbeck2012free, gallego2013full, ramanathan2016randomness, brandao2016realistic, ramanathan2018practical, kessler2020device}. 

As a model of an imperfect seed, one may consider the $\varepsilon$-SV source \cite{santha1986generating}, a model of a biased coin where the individual coin tosses are not independent but rather the bits $R_i$ produced by the source obey 
    $\frac{1}{2} - \varepsilon \leq P(R_i=0 | R_{i-1}, \dots, R_1, W) \leq \frac{1}{2} + \varepsilon$.
The parameter $0 \leq \varepsilon < \frac{1}{2}$ described the reliability of the source, with $\varepsilon = 0$ being the ideal random seed and $W$ denotes any side information, possibly held by an adversary. It is worth remarking that more general `min-entropy' sources are also possible, wherein only a lower bound on the min-entropy (the negative logarithm of the maximum probability of any output string) produced by the source is assumed. 

With regards to the task of self-testing, an important result was the formulation of a general class of Bell inequalities known as the tilted-CHSH inequalities suitable for self-testing general pure two-qubit entangled states. Specifically, in the simplest bipartite Bell scenario with two binary observables $A_0, A_1$ for Alice and two binary observables $B_0, B_1$ for Bob, the following family of tilted CHSH operators was introduced in \cite{acin2012randomness}
\begin{equation}
T_{\beta} = \beta A_0 + A_0 B_0 + A_0 B_1 + A_1 B_0 - A_1 B_1,
\end{equation}
where $\beta \in [0,2)$ is a parameter with $\beta = 0$ corresponding to the well-known CHSH operator. The maximum value of this tilted CHSH quantity in classical theories is easily seen to be $2+ \beta$. In \cite{bamps2015sum}, it was shown that the optimal quantum value of $\textcolor{black}{T_{\beta}}^{\text{max}} = \sqrt{8 + 2 \beta^2}$ can only be achieved when specific reference observables $A_x, B_y$ are measured on the state $|\psi_{\theta} \rangle = \cos{\theta} | 00 \rangle + \sin{\theta} |11 \rangle$ with $\theta = \frac{1}{2} \arctan{\sqrt{\frac{2}{\beta^2} - \frac{1}{2}}}$. Specifically, it was shown that the observation of the expectation value $\langle \textcolor{black}{T_{\beta}} \rangle \geq \textcolor{black}{T_{\beta}}^{\max} - e$ for the tilted CHSH operator by measuring a physical state $|\psi' \rangle$ with observables $A'_x$ and $B'_y$ implies the existence of an isometry $\Phi = \Phi_A \otimes \Phi_B$ and a state $|\text{junk} \rangle$ such that $\| \Phi\left(A'_x \otimes B'_y |\psi' \rangle \right) - |\text{junk} \rangle \otimes (A_x \otimes B_y) | \psi \rangle \| \leq e'$, where $x, y \in \{-1,0,1\}$ (with subscript $-1$ indicating the identity operator) and where $e' = O\left(\sqrt{e} \right)$. The family of tilted CHSH inequalities has since played a crucial role in several aspects, including (a) showing the quantitative inequivalence between the amount of non-locality and the amount of certified randomness \cite{acin2012randomness}, (b) being the building block that enabled the self-testing of all bipartite pure (two-qudit) entangled states \cite{coladangelo2017all}, (c) enabling the formulation of a device-independent quantum random number generation (DIQRNG) protocol that only requires a sublinear amount of quantum communication \cite{bamps2018device} \textcolor{black}{(i.e., produces $n$ bits of output randomness with a total of $ \Omega(n^k \log n)$ ebits where $\frac{7}{8} < k < 1$),}
and (d) unbounded randomness certification from a single pair of entangled qubits using a sequence of measurements \cite{curchod2017unbounded}.

A natural question is whether a corresponding family of inequalities can be formulated in the scenario of (arbitrarily) limited measurement independence so that the above (and further such DI) results can be achieved in the setting when the parties are not assumed to possess perfect random seeds. In section~\ref{tilted_hardy}, we answer this question in the positive by formulating a class of tilted Hardy paradoxes that allow to self-test general pure two-qubit entangled states (except the maximally entangled state). As shown in \cite{putz2014arbitrarily}, tests of Hardy paradoxes (in an equivalent formulation as `measurement-dependent' locality inequalities) allow for arbitrary small measurement independence making them ideal candidates for device-independent tasks when only weak seeds are available. In this section~\ref{tilted_hardy}, we also derive expressions for the amount of randomness that can be certified from the maximum violation of the tilted Hardy tests in terms of the \textit{guessing probability} by an adversary holding a quantum system that is potentially correlated to the devices involved in the test. In section~\ref{noise_cases}, we compute the guessing probability in the noisy scenario of non-maximal violation, here distinguishing between two cases: (a) a scenario of colored noise where the `zero' constraints in the tilted Hardy paradox are satisfied but the non-zero Hardy probability is non-maximal, and (b) a scenario of white noise where we consider the non-maximal violation of a Bell expression derived from the tilted Hardy paradox. While an amount of local randomness up to the maximum possible value of $1$ bit can be certified by the tilted Hardy tests, the amount of global randomness is limited to a value of approximately $1.6806$ bits. Nevertheless, we show that the derived results present an improvement in generation rate over the state-of-the-art protocols of randomness amplification against quantum adversaries. Meanwhile, we slightly change the coefficients of the expression of Hardy probability and show that the derived Bell expression provides a robust self-testing statement of the corresponding optimal quantum system. In section~\ref{nosig}, we derive the analytical expression for the guessing probability in the scenario of an adversary that is allowed to prepare bipartite devices for the honest parties constrained only by the no-signaling principle, a result that also finds application in the state-of-the-art protocols for randomness amplification against no-signaling adversaries \cite{ramanathan2018practical}. Finally, in section~\ref{gadgetsec}, we present a class of Hardy paradoxes with more inputs and outputs that potentially allow to certify the maximal amount of global randomness of $2 \log d$ bits for dimensions $d = 4, 8$. To do this, we exploit a recently discovered connection between Hardy paradoxes and substructures of Kochen Specker proofs termed $01$-gadgets \cite{ramanathan2022large}.  We conclude with a discussion on future research directions.

\section{A family of Tilted Hardy paradoxes: Self-testing general pure two-qubit entangled states}\label{tilted_hardy}

We work in the simplest bipartite Bell scenario with Alice and Bob performing measurements corresponding to two binary observables each. Specifically, Alice and Bob provide inputs $x\in\{0,1\}$ and $y\in\{0,1\}$ to their respective device, and obtain respective binary outcomes $a\in\{0,1\}$ and $b\in\{0,1\}$. The behavior of the device is described by a set of probability distributions $\{P_{A,B|X,Y}(a,b|A_x,B_y)\}$ where $P_{A,B|X,Y}(a,b|A_x,B_y)$ denotes the probability to obtain outputs $a, b$ conditioned on measurement inputs $x, y$ for Alice and Bob. We define a family of tilted Hardy tests in this scenario as given by the following set of constraints: 

\begin{align}\label{eq:w-Hardy}
&P_{A,B|X,Y}(a,b|A_x,B_y)=0, \; \forall \; (a,b|x,y)\in S_{Hz} :=\{(0,1|0,1),\; (1,0|1,0), \; (0,0|1,1)\} \nonumber \\
&P^{w}_{Hardy} := P_{A,B|X,Y}(0,0|A_0,B_0)+w\cdot P_{A,B|X,Y}(1,1|A_0,B_0) > \max(0,w)
\end{align}
where $w\in(-\frac{1}{4},1)$ is a parameter, and $\max(0,w)$ is the maximum value of the Hardy probability $P^w_{Hardy}$ in any classical local hidden variable (LHV) theory. This latter bound can be seen from the fact that the enforcement of the three Hardy `zero' constraints $P_{A,B|X,Y}(a,b|A_x,B_y)=0$ for $(a,b|x,y) \in S_{Hz}$ implies as in the original Hardy paradox that $P_{A,B|X,Y}(0,0|A_0,B_0) = 0$ and in any LHV model $P_{A,B|X,Y}(1,1|A_0,B_0) \leq 1$. As we shall see, quantum theory allows for a value of $P^{w}_{Hardy}$ larger than this classical bound. Indeed when $w=0$, we recover the original Hardy paradox \cite{hardy1993nonlocality} for which it is well-known that $P_{Hardy}^{w=0}=P_{A,B|X,Y}(0,0|A_0,B_0)$ can achieve a maximal value of $\frac{5\sqrt{5}-11}{2}$ in quantum theory. Furthermore, as shown in \cite{rabelo2012device}, this quantum behavior is achieved by a unique two-qubit state and measurements up to local isometries.





As anticipated in the Introduction, the motivation for introducing the tilted Hardy paradoxes is that as shown in \cite{putz2014arbitrarily}, Hardy paradoxes can be used to certify quantum non-locality even in scenarios when the measurement settings are not chosen completely freely. Specifically, in any scenario where the parties choose their measurement settings using weak sources of randomness that obey $l\leq P_{X,Y}(x,y|\lambda)\leq h$ for $l,h\in(0,1)$, the tilted Hardy test in Eq. \eqref{eq:w-Hardy} can be converted into a Measurement-Dependent Locality (MDL) inequality as:
\begin{align}\label{eq:MDL}
    &&l\cdot \left[ P_{Hardy}^{w}-\max(0,w) \right]P_{X,Y}(A_0,B_0)-h\cdot [P_{X,Y}(A_0,B_1)P_{A,B|X,Y}(0,1|A_0,B_1) \nonumber \\&&+ P_{X,Y}(A_1,B_0)P_{A,B|X,Y}(1,0|A_1,B_0)+P_{X,Y}(A_1,B_1)P_{A,B|X,Y}(0,0|A_1,B_1)] \leq 0,
\end{align}
where $P_{X,Y}(A_x,B_y)$ denotes the probability of choosing inputs $x,y$. Furthermore, as the following Prop.~\ref{prop:w-hardy} shows (we defer its proof to Appendix~\ref{proof-w-hardy}), the tilted Hardy tests can be used to self-test every entangled pure two-qubit state other than the maximally entangled state. 

\begin{proposition}\label{prop:w-hardy}
Suppose that the Hardy zero constraints  $P_{A,B|X,Y}(0,1|A_0,B_1)=$ $\qquad$ $ P_{A,B|X,Y}(1,0|A_1,B_0)=P_{A,B|X,Y}(0,0|A_1,B_1)=0$ are satisfied. Then in any local hidden variable theory it holds that $P_{Hardy}^{w,LHV}\leq\max(0,w)$. On the other hand, quantum theory allows to achieve $P_{Hardy}^{w,Q}=\frac{(4w+5)\sqrt{4w+5}-(12w+11)}{2(w+1)} > P_{Hardy}^{w,LHV}$ for $w \in (-\frac{1}{4},1)$. Furthermore, up to local isometries, $P_{Hardy}^{w,Q}$ is uniquely achieved by the two-qubit state $|\psi_{\theta}\rangle=\cos(\frac{\theta}{2})|00\rangle-\sin(\frac{\theta}{2})|11\rangle$ with $\theta=\arcsin{(3-\sqrt{4w+5})}$ and the measurements $\{A_0,A_1\}$, $\{B_0,B_1\}$ given as
\begin{align}\label{eq:w-meas}
   A_0=B_0&=\frac{-(2+\sin\theta)(1-\sin\theta)^{\frac{1}{2}}}{(2-\sin\theta)(1+\sin\theta)^{\frac{1}{2}}}\cdot Z+\frac{-\sqrt2\sin\theta(\sin\theta)^{\frac{1}{2}}}{(2-\sin\theta)(1+\sin\theta)^{\frac{1}{2}}}\cdot X \nonumber, \\
  A_1=B_1&=\frac{-(1-\sin\theta)^{\frac{1}{2}}}{(1+\sin\theta)^{\frac{1}{2}}}\cdot Z+\frac{\sqrt2(\sin\theta)^{\frac{1}{2}}}{(1+\sin\theta)^{\frac{1}{2}}}\cdot X ,
\end{align}
where $X$ and $Z$ are the usual Pauli matrices.
\end{proposition}

Let us briefly observe that no loss of generality is accrued in considering the expression $P_{Hardy}^{w}$ in Eq. \eqref{eq:w-Hardy} - when the three Hardy zero constraints are satisfied, $P_{Hardy}^{w}$ constitutes a general (non-zero) Hardy expression, i.e., no additional gain is obtained by considering the quantum value of an expression such as 

\begin{equation}
\label{general_Hardy}
\begin{split}
  P_{Hardy}=&\alpha_{0,0} P_{A,B|X,Y}(0,0|A_0,B_0) + \alpha_{0,1} P_{A,B|X,Y}(0,1|A_0,B_0) \\&+ \alpha_{1,0} P_{A,B|X,Y}(1,0|A_0,B_0) + \alpha_{1,1} P_{A,B|X,Y}(1,1|A_0,B_0)
  \end{split}
\end{equation} 
for arbitrary coefficients $\alpha_{i,j}$ with $i,j \in \{0,1\}$. To see this, consider an arbitrary bipartite pure state $|\psi\rangle=\sum _k \lambda_k|k_A\rangle|k_B\rangle$, where $\{|k_A\rangle\}$and $\{|k_B\rangle\}$ are basis sets of Hilbert spaces $\mathbf{H}_A$ and $\mathbf{H}_B$ respectively and $\lambda_k\geq0$ are the corresponding Schmidt coefficients. First observe that Jordan's lemma allows for a reduction to the setting  where Alice and Bob hold qubit systems, i.e., $\mathbf{H}_A=\mathbf{H}_B=\mathbb{C}^2$. Furthermore, when the Hardy zero constraints are satisfied, as shown in Appendix~\ref{proof-w-hardy}, Alice and Bob's measurement observables are identical, i.e., $A_0=B_0$ and $A_1=B_1$ giving rise to the symmetry $P(0,1|A_0,B_0)=P(1,0|A_0,B_0)$.
Under these conditions, the arbitrary Hardy expression
    $P_{{Hardy}}= \sum_{i,j = 0,1} \alpha_{i,j}  P_{A,B|X,Y}(i,j|A_0,B_0)$
reduces to
\begin{equation}
\begin{split}
P_{{Hardy}}=&\frac{\alpha_{0,1}+\alpha_{1,0}}{2}+\left(\alpha_{0,0}-\frac{\alpha_{0,1}+\alpha_{1,0}}{2} \right)P_{A,B|X,Y}(0,0|A_0,B_0)\\
&+\left(\alpha_{1,1}-\frac{\alpha_{0,1}+\alpha_{1,0}}{2} \right)P_{A,B|X,Y}(1,1|A_0,B_0).
\end{split}
\end{equation}
Removing the offset term $\frac{\alpha_{0,1}+\alpha_{1,0}}{2}$ and denoting the ratio of $\left(\alpha_{1,1}-\frac{\alpha_{0,1}+\alpha_{1,0}}{2} \right)$ and \\ $\left(\alpha_{0,0}-\frac{\alpha_{0,1}+\alpha_{1,0}}{2} \right)$ as $w$ reduces the general Hardy expression to $P_{Hardy}^{w}$.

The fact that the maximum violation of the tilted Hardy tests is realised by self-testing quantum behaviors implies that one may apply them in schemes to generate certified randomness starting from arbitrarily weak seeds (we elaborate on this in the coming sections). We may quantify the amount of global randomness obtained by the measurement of observables $A_x$ and $B_y$ on a state $|\psi \rangle \in \mathbf{H}_A \otimes \mathbf{H}_B$ through the guessing probability $G_{\psi, A_x, B_y} = \max_{a,b} P^{\psi}_{A,B|X,Y}(a,b|A_x, B_y)$ where $P^{\psi}_{A,B|X,Y}(a,b|A_x, B_y)$ denotes the outcome probabilities obtained by measurements of $A_x, B_y$ on $|\psi \rangle$. The corresponding amount of global randomness (in bits) is then given by the min-entropy $H_{\infty}(|\psi \rangle, A_x, B_y) = - \log_2 G_{\psi, A_x, B_y}$. Similarly, one can quantify the amount of local randomness through the guessing probability of marginal outcomes of each local party.
One may then define the maximum global randomness certifiable from any input pair $x,y$ the tilted Hardy tests for any given $w$ as \textcolor{black}{$H^{\text{global}}_{\infty} = \max_{x, y} H_{\infty}(|\psi_{\theta} \rangle, A_x, B_y)$} where $|\psi_{\theta}\rangle, A_x, B_y$ are given by the self-testing quantum behavior from Prop.~\ref{prop:w-hardy} with $\theta = \arcsin{(3-\sqrt{4w+5})}$. Analogously, one can quantify the amount of local randomness obtained by the measurement of $A_x$ or $B_y$, the symmetry of the measurement observables of Alice and Bob means that we obtain \textcolor{black}{$H^{\text{local}}_{\infty} = \max_x H_{\infty}(|\psi_{\theta}\rangle, A_x) = \max_y H_{\infty}(|\psi_{\theta}\rangle, B_y)$}.  \textcolor{black}{As the pure state $|\psi_{\theta}\rangle$ is uncorrelated with any third party’s system, the maximal local and global randomness is against quantum eavesdroppers.}

We calculate the maximum amount of certifiable local and global randomness from the tilted Hardy tests as: 
\begin{equation}\label{local_rand_self}
\begin{split}
\textcolor{black}{H^{\text{local}}_{\infty}}&=H_{\infty}(|\psi_{\theta}\rangle, A_1)=H_{\infty}(|\psi_{\theta}\rangle, B_1)
=1-\log _{2}\left[1+\cos\theta \frac{1-\tan{(\theta/2)}}{1 + \tan{(\theta/2)}} \right]
\end{split}
\end{equation}
where $\theta = \arcsin{(3-\sqrt{4w+5})}$ from Prop.~\ref{prop:w-hardy}, and

\begin{align}\label{eq:max-global}
\textcolor{black}{H^{\text{global}}_{\infty}}=\left\{
\begin{array}{ll}
1-\log _{2}[\frac{(3-\sqrt{4w+5})^3}{(\sqrt{4w+5}-1)^2}] &\quad{-\frac{1}{4}<w\leq w_0}\\
1-\log _{2}[\frac{8\sqrt{4w+5}-16}{(\sqrt{4w+5}-1)^2}] &\quad{w_0<w\leq w_1}\\
1-\log _{2}[3-\sqrt{4w+5}] &\quad{w_1<w\leq \frac{1}{9}}\\
1-\log _{2}[2\sqrt{4w+5}-4] &\quad{\frac{1}{9}<w < 1}\\
\end{array} \right.
\end{align}
where $w_0=\frac{1}{4}[-5+(3+4(\frac{2}{3(9+\sqrt{177})})^{\frac{1}{3}}-(4+\frac{4\sqrt{177}}{9})^{\frac{1}{3}})^2]\approx-0.1546$ and 
$w_1=\frac{1}{4}[-5+(3+\frac{1}{3}(-4+10(\frac{2}{-11+3\sqrt{69}})^{\frac{1}{3}}-(12\sqrt{69}-44)^{\frac{1}{3}}))^2]\approx-0.1054$.

Note that the reason why \textcolor{black}{$H^{\text{global}}_{\infty}$} is a piecewise function is that the maximum global randomness is achieved by different measurement inputs $x,y$ for different values of $w$. 
We see that the tilted Hardy tests allow to certify up to $1$ bit of local randomness in the limit as $w \rightarrow -\frac{1}{4}$ (as the state tends to a maximally entangled state). 
Similarly, one can obtain up to $1.6806$ bits of global randomness for $w = w_0$. In \cite{acin2012randomness}, it was observed that the amount of certified randomness is an inequivalent resource to non-locality and entanglement, in that high global randomness (close to $2$ bits) may be certified by the tilted CHSH inequality even when using states that are almost unentangled, with the corresponding quantum behaviors being close to the set of local hidden variable behaviors. A similar phenomenon can also be observed in the expression for the global randomness in Eq. \eqref{eq:max-global}. While the maximum global randomness is achieved at $w \approx -0.1546$, the largest separation between the quantum and the LHV set occurs at $w = 0$ (when $\sin{\theta} = 3 - \sqrt{5}$), and the optimal state is close to maximally entangled as $w \rightarrow -\frac{1}{4}$ (when $\sin{\theta} \rightarrow 1$).

\textcolor{black}{Note that the above self-testing result of Proposition.~\ref{prop:w-hardy} does not consider an SV-source input distribution per se, since the biased SV-source themselves only give bounds of each input probability $l\leq P_{X,Y}(x,y|\lambda)\leq h$,
and in this case it is not possible to even determine the exact quantum value of the corresponding measurement-dependent-locality (MDL) inequality. However, if the input distribution is explicitly known, the tilted Hardy paradoxes can be modified into the MDL version of the tilted Hardy tests and then the corresponding self-testing statement can be derived correspondingly, such that every pure two-qubit entangled state (other than the maximally entangled state) can be self-tested under arbitrary limited of measurement dependence.}

\section{Applying the Tilted Hardy Test in device-independent randomness generation against Quantum adversaries}\label{noise_cases}
In the previous section, we have derived the optimal amount of randomness that can be certified when the maximal violation of the tilted Hardy test is observed for any value of the parameter $w\in(-\frac{1}{4},1)$. When using the Hardy paradox in an actual device-independent protocol to certify randomness against a quantum adversary (such as \cite{li2015device}), it is important to consider the noisy scenario of non-maximal violation.
Specifically, we consider the three-party scenario of Alice, Bob and Eve holding a tripartite quantum state $\rho_{ABE}$ belonging to the Hilbert spaces $\mathbf{H}_A\otimes\mathbf{H}_B\otimes\mathbf{H}_E$, i.e., Eve holds a quantum system that may be entangled with that held by Alice and Bob. Alice and Bob perform measurements chosen using some $\epsilon$-SV source, observe a certain value in the Hardy test and attempt to generate randomness from their measurement outcomes for some specific setting $(A_x,B_y)$. Eve performs a single measurement with outcome $e\in\{0,1\}^{2}$ which will constitute her guess of the measurement results of Alice and Bob for this setting.

Formally, Eve's guessing probability of Alice and Bob's outputs can be written as the maximization of
\begin{equation}
    {P_{guess}(A,B|A_x,B_y,E)}=\sum_{a,b\in\{0,1\}}P_{A,B,E|X,Y}(a,b,(e=a,b)|A_x,B_y)
\end{equation}
subject to some given constraints. 
Here $P_{A,B,E|X,Y}(a,b,e|A_x,B_y)= \linebreak \Tr\left[ A_{a|x} \otimes B_{b|y} \otimes \Pi_e \rho_{ABE} \right]$
where  $A_{a|x}$ and $B_{b|y}$ are the projectors corresponding to the outcomes $a$ and $b$ for Alice and Bob respectively, and $\Pi_e$ denotes the projector corresponding to Eve's measurement outcome $e$. The min-entropy is denoted as $ H_{\infty}(A,B|A_x,B_y,E)=-\log_2P_{guess}(A,B|A_x,B_y,E)$. A similar expression holds when Eve attempts to guess the local outcomes of one of the parties alone.

We consider two noise models that we term (1) colored noise and (2) white noise - in the former, we consider the scenario where the observed behavior $\{P_{A,B|X,Y}\}$ stays on the boundary of the quantum set (i.e., the three Hardy zero constraints are satisfied, while the non-zero Hardy probability takes on some intermediate value between the local and the quantum maximum), while in the latter, we consider the scenario where the observed behavior $\{P_{A,B|X,Y}\}$ is in the interior of the quantum set (i.e., the honest parties simply observe some non-optimal value for a Bell expression derived from the Hardy test).  

\begin{enumerate}
    \item Colored Noise: In this case, we consider that Alice and Bob's reduced system $\rho_{AB}$ satisfies the three Hardy zero constraints, while the $P_{Hardy}^{w}$ is non-maximal, that is $P_{Hardy}^{w}$ lies in the intermediate  range between $\max(0,w)$ and $P_{Hardy}^{w,Q}$.
    
    \item White Noise: In this case, we slightly change the coefficients of Hardy probability in Eq.~\eqref{general_Hardy} to $\alpha_{0,0}=1, \alpha_{0,1}=\alpha_{1,0}=\frac{1-\alpha}{2}, \alpha_{1,1}=0$, and consider that Alice and Bob's reduced system $\rho_{AB}$ achieves some (non-maximal) value of the corresponding derived Bell expression (by subtracting the Hardy zero constraints from the non-zero Hardy probability):
    \begin{equation}\label{eq:I-Hardy}
    \begin{split}
        I_{\alpha}^{Hardy}=&[P_{A,B|X,Y}(0,0|A_0,B_0)+\frac{1-\alpha}{2}P_{A,B|X,Y}(0,1|A_0,B_0)\\
        &+\frac{1-\alpha}{2}P_{A,B|X,Y}(1,0|A_0,B_0)-\max(0,\frac{1-\alpha}{2})]-[P_{A,B|X,Y}(0,1|A_0,B_1)\\
        &+P_{A,B|X,Y}(1,0|A_1,B_0)+P_{A,B|X,Y}(0,0|A_1,B_1)].
    \end{split}
    \end{equation}
It is worth noting that a similar expression for $\alpha=1$ (in which case we recover the original Hardy paradox and $I_{\alpha=1}^{Hardy}$ corresponds to the CHSH inequality) has found application in the state-of-the-art device-independent randomness amplification protocol against an adversary holding quantum side information\textcolor{black}{~\cite{kessler2020device}}. 

\end{enumerate}


\subsection{Colored Noise}
In this subsection, we show the guessing probability for the value of $w$ that results in the largest min-entropy in the noiseless setting, i.e., we set $w = w_0$ (in this case we note that the randomness is certified from the input $(A_0, B_0)$ ). Suppose that Alice and Bob observe a value of the Hardy probability $P^{w}_{Hardy} =: p$ that lies between the classical value \textcolor{black}{$\max(0,w_0)=0$} and the quantum maximum $P^{w = w_0, Q}_{Hardy} = \frac{(4 w_0+5)\sqrt{4 w_0+5}-(12 w_0 +11)}{2( w_0 +1)}$.   
We compute the analytical bound of the guessing probability as a function of $p$ for the input $(A_0, B_0)$ (the detailed computation is deferred to Appendix~\ref{randomness under quantum adversary}) as 
\begin{equation}
\begin{split}
&\textcolor{black}{P_{guess}(A,B|A_0,B_0,E)}=\left\{
\begin{array}{ll}
\frac{\textcolor{black}{(\left[P_{A,B|X,Y}(0,1|A_0,B_0)\right]_{\theta_{2}^{\textcolor{black}{\widetilde{p^{*}}},w_0}})}-1}{\textcolor{black}{\widetilde{p^{*}}}}\cdot p+1, \quad & 0\leq p\leq p^{*},\\
\left[ P_{A,B|X,Y}(1,1|A_0,B_0) \right]_{\theta_1^{p,w_0}}, \quad & p^{*}< p\leq P_{Hardy}^{w_0,Q},\\
\end{array} \right.\\
\end{split}
\end{equation}  
where $\textcolor{black}{p^{*}} \approx 0.01366$, $\textcolor{black}{\widetilde{p^{*}}} \approx 0.01563$, $\theta_i^{p,w_0}$ for $i=1,2$ are the two solutions of  Eq.~\eqref{hardy_value_vio} for $P_{Hardy}^{w_0} = p$ when the three zero constraints are satisfied.

\subsection{White Noise}\label{section_new_Bell}
\begin{figure}[ht]
  \centering
  \includegraphics[width=0.75\textwidth]{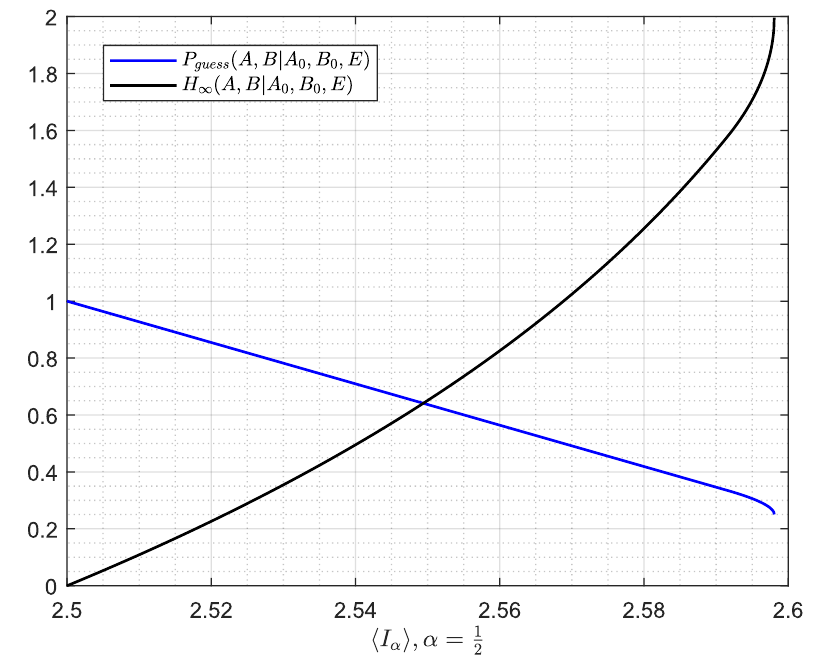}
  \caption{At the optimal point, $I_{\alpha=1/2}$, the inequality allows to certify up to $\textcolor{black}{H_{\infty}(A,B|A_0,B_0,E)}= 2$ bits global randomness. The guessing probability $\textcolor{black}{P_{guess}(A,B|A_0,B_0,E)}$ and corresponding Min-entropy $\textcolor{black}{H_{\infty}(A,B|A_0,B_0,E)}$ (bits per round) versus $I_{\alpha=1/2}$ calculated at the second level of the NPA hierarchy is shown.  }\label{min_entrpy_eavesdropping}
\end{figure}

In this subsection, we consider the scenario in which Alice and Bob's system $\rho_{AB}$ achieves a non-maximal value of the Bell expression ${I}_{\alpha}$ derived from the tilted Hardy test. To simplify matters, we formulate this Bell expression in terms of the correlators $\langle A_x B_y \rangle = \sum_{a,b=0,1} (-1)^{a + b} P_{A,B|X,Y}(a,b|A_x,B_y)$, $\langle A_x \rangle = \sum_{a=0,1} (-1)^a P_{A|X}(a|A_x)$ (and similar for $\langle B_y \rangle$). To this end, let us subtract the three zero terms from the Hardy non-zero probability \eqref{general_Hardy} (with coefficients $\alpha_{0,0}=1, \alpha_{0,1}=\alpha_{1,0}=\frac{1-\alpha}{2}, \alpha_{1,1}=0$): 
\begin{equation}\label{Bell_ineq1}
\begin{split}
        I_{\alpha}^{Hardy}=&[P_{A,B|X,Y}(0,0|A_0,B_0)+\frac{1-\alpha}{2}P_{A,B|X,Y}(0,1|A_0,B_0)+\frac{1-\alpha}{2}P_{A,B|X,Y}(1,0|A_0,B_0)\\&-\max(0,\frac{1-\alpha}{2})]-\textcolor{black}{[P_{A,B|X,Y}(0,1|A_0,B_1)+P_{A,B|X,Y}(1,0|A_1,B_0)+P_{A,B|X,Y}(0,0|A_1,B_1)]}\\
        =&\frac{\alpha \langle A_0B_0\rangle+\langle A_0B_1\rangle+\langle A_1B_0\rangle-\langle A_1B_1\rangle+(-\alpha-1)}{4}-\max(0,\frac{1-\alpha}{2})\\
        =&\frac{1}{4}[\langle I_{\alpha}\rangle+(-\alpha-1)]-\max(0,\frac{1-\alpha}{2}),
\end{split}
\end{equation}
where $I_{\alpha}:=\alpha A_0B_0+A_0B_1+A_1B_0-A_1B_1$, which is termed as $\alpha$-CHSH expression in Appendix~\ref{robust_self_testing_App}. And in Appendix~\ref{robust_self_testing_App}, We analytically compute the maximum values of this Bell expression $I_\alpha$ in quantum and classical theories, and show it provides a robust self-testing statement for the optimal quantum strategy.
With the optimal quantum strategy, one can obtain $\textcolor{black}{H^{global}_{\infty}}=2$ bits of global randomness when $\alpha=\frac{1}{2}$ \textcolor{black}{(See Fig.~\ref{min_entrpy_eavesdropping} for details on the relation between the global randomness and violation of the inequality $I_{\alpha=\frac{1}{2}}\leq \frac{5}{2}$)}. 
It is in the form of $I_{\alpha}$ that Hardy tests find application in testing non-locality with arbitrarily limited measurement independence, using the so-called measurement-dependent locality (MDL) inequalities \textcolor{black}{(see following Eq.~\eqref{MDL} for our case)}\cite{putz2014arbitrarily}. 

To elaborate, consider the scenario in which the measurement settings $(x,y)$ are chosen using a weak seed such as an $\epsilon$-SV source which obeys
$l\leq P_{X,Y}(x,y|\lambda)\leq h$,
where $\lambda$ denotes any side information that may be held by an adversary about the source. Here $0 < l \leq h < 1$, and for the specific case of the $\epsilon$-SV source, $l = (\frac{1}{2} - \epsilon)^2$ and $h = (\frac{1}{2} + \epsilon)^2$. The expression $I_{\alpha}$ gives rise to the following tilted measurement-dependent-locality expression
\begin{equation}\label{MDL}
\begin{split}
I^{MDL}_{\alpha}=&l[P_{A,B,X,Y}(0,0,A_0,B_0)+\frac{1-\alpha}{2}P_{A,B,X,Y}(0,1,A_0,B_0)\\
&+\frac{1-\alpha}{2}P_{A,B,X,Y}(1,0,A_0,B_0)-\max(0,\frac{1-\alpha}{2})P_{X,Y}(A_0,B_0)]\\
&-h[P_{A,B,X,Y}(0,1,A_0,B_1)+P_{A,B,X,Y}(1,0,A_1,B_0)+P_{A,B,X,Y}(0,0,A_1,B_1)]\\
        \leq & lh [P_{A,B|X,Y}(0,0|A_0,B_0)+\frac{1-\alpha}{2}P_{A,B|X,Y}(0,1|A_0,B_0)\\
        &+\frac{1-\alpha}{2}P_{A,B|X,Y}(1,0|A_0,B_0)-\max(0,\frac{1-\alpha}{2})]\\
        &-hl [P_{A,B|X,Y}(0,1|A_0,B_1)+P_{A,B|X,Y}(1,0|A_1,B_0)+P_{A,B|X,Y}(0,0|A_1,B_1)]\\
        &\textcolor{black}{=lh\cdot I_{\alpha}^{Hardy} = \frac{lh}{4}[\langle I_{\alpha}\rangle+(-\alpha-1)]-lh \cdot\max(0,\frac{1-\alpha}{2})}
\end{split}
\end{equation}
with $\alpha\in(1/3,+\infty)$. The corresponding classical bound is readily seen to be $I^{MDL}_{\alpha}\overset{{LHV}}{\leq} 0$, and the quantum bound is directly obtained from the optimal quantum strategy for the Bell expression $I_{\alpha}$ in Eq.~\eqref{Bell_ineq1}. 
\color{black} From Eq.~\eqref{MDL} we get:
\begin{equation}\label{mdl-Ialpha}
\begin{split}
        I_{\alpha}^{Hardy}&\geq \frac{I_{\alpha}^{MDL}}{lh},\\
        \langle I_{\alpha}\rangle&\geq 4\left(\frac{I_{\alpha}^{MDL}}{lh} +\max(0,\frac{1-\alpha}{2})\right)+(\alpha+1) .   \\
\end{split}
\end{equation}
By numerically calculating the guessing probability  $P_{\text{guess}}(A_0,B_0|E)$ and corresponding min-entropy $H_{\infty}(A,B|A_0,B_0,E)$ versus the $\langle I_{\alpha}\rangle$ using the NPA hierarchy (of level 2), one can then use them with the relation in Eq.~\eqref{mdl-Ialpha} to bound the min-entropy versus  $ I_{\alpha}^{MDL}$ in the MDL scenarios. We show the three cases for $\alpha=\frac{1}{2},\alpha=\frac{4}{5}$ and $\alpha=1$ in Fig.~\ref{mdl-vio}, from which and the quantum-classical value gap we calculate in Appendix~\ref{robust_self_testing_App}, one can see there is a trade-off between the maximal randomness one can certify from the MDL inequality $I_{\alpha}^{MDL}$ and the robustness of this MDL inequality $I_{\alpha}^{MDL}$. $I_{\alpha=\frac{1}{2}}^{MDL}$ gives the highest randomness (2 bits) but it is not very robust, and $I_{\alpha=1}^{MDL}$ is the most robust one in this whole family of MDL inequality. $I_{\alpha=\frac{4}{5}}^{MDL}$ is one choice that balances the randomness and robustness between these two extreme cases.

\begin{figure}[H]
\centering
\subfigure[]{\hspace{-2cm}
\begin{minipage}[t]{1.17\linewidth}
\centering
\includegraphics[width=1\textwidth]{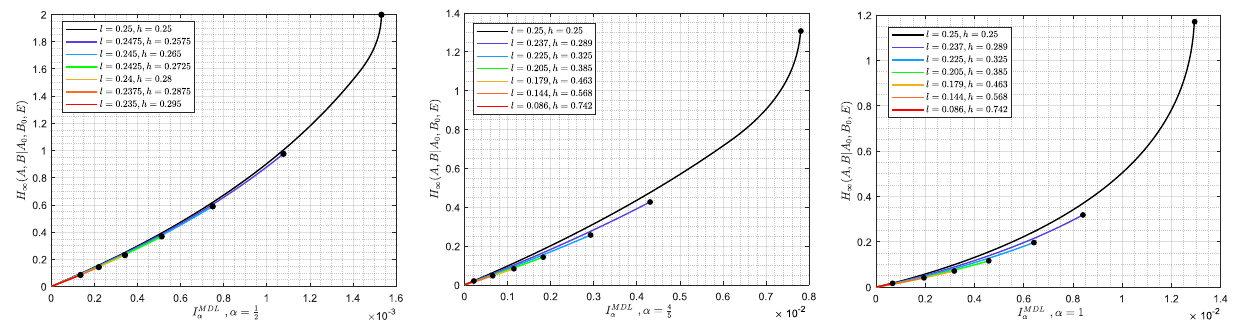}
\label{mdl-vio}
\end{minipage}%
}%
\\
\subfigure[]{
\begin{minipage}[t]{0.9\linewidth}
\centering
\includegraphics[width=1\textwidth]{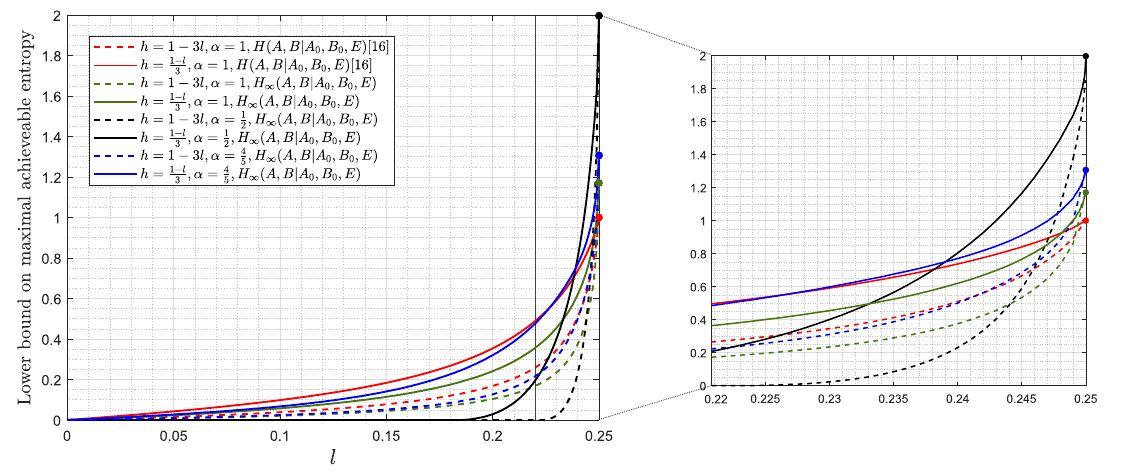}
\label{mdl-l}
\end{minipage}%
}
\centering
  \caption{\textcolor{black}{(a) Min-entropy bound in the MDL scenario verses $I_{\alpha}^{MDL}$ for $\alpha=\frac{1}{2},\alpha=\frac{4}{5},\alpha=1$ with the given lower and higher bounds $l,h$ of a weak SV-source. The black dots denote the lower bound on the maximal possible quantum value of $I_{\alpha}^{MDL}$ with the given lower and higher bounds $l,h$ of the SV-source. (b) Lower bound on the maximal achievable entropy of $I_{\alpha}^{MDL}$ ($\alpha=\frac{1}{2},\alpha=\frac{4}{5},\alpha=1$) versus the lower bound $l$ of a weak SV-source. Here for the weak SV-source with fixed $l$, we consider two extreme cases of the input distribution (1) three settings are chosen with probability $h$ and the remaining one is $l$, i.e. $l=1-3h$ ($h = \frac{1-l}{3}$), and (2) three settings are chosen with probability $l$ and the remaining one is $h$, i.e.  $h = 1-3l$. As a comparison, we also plot the results in \cite{kessler2020device}, in which the authors calculated the von Neumann entropy in the MDL scenario for $\alpha=1$.
  }}\label{MDL_min_entrpy_eavesdropping}
\end{figure}
Since only the lower and higher bounds $l,h$ of each input are known instead of the explicitly input distribution, the MDL inequality $I_{\alpha}^{MDL}$ is not fixed, thus the maximal quantum value cannot be calculated. Nevertheless one can derive a lower bound of the maximal quantum value of $I_{\alpha}^{MDL}$ by considering the worst-case input distribution, i.e. $P_{X,Y}(A_0,B_0) = l$, and $P_{X,Y}(A_x,B_y) = h$ for $(x,y) \neq (0,0)$, then we get:
\begin{equation}
\begin{split}
I_{\alpha}^{MDL} & \geq l^2 [P_{A,B|X,Y}(0,0|A_0,B_0)+\frac{1-\alpha}{2}P_{A,B|X,Y}(0,1|A_0,B_0)\\
&+\frac{1-\alpha}{2}P_{A,B|X,Y}(1,0|A_0,B_0)-\max(0,\frac{1-\alpha}{2}) ] - h^2 [P_{A,B|X,Y}(0,1|A_0,B_1) \\
&+ P_{A,B|X,Y}(1,0|A_1,B_0) + P_{A,B|X,Y}(0,0|A_1,B_1) ] =: \tilde{I}_{\alpha}^{MDL}
\end{split}
\end{equation}
We use the NPA hierarchy of level 2 to numerically calculate the maximal quantum value of $\tilde{I}_{\alpha}^{MDL}$ which is the lower bound of the maximal quantum value of $I_{\alpha}^{MDL}$. In Fig.~\ref{mdl-l}, we then use this lower bound of $I_{\alpha}^{MDL}$ to show the lower bound on the maximal achievable min-entropy of $I_{\alpha}^{MDL}$ for $\alpha=\frac{1}{2},\alpha=\frac{4}{5},\alpha=1$ versus the lower bound $l$ of the SV-source, in which for a fixed $l$, we consider two extreme cases of the input distribution (1) three settings are chosen with probability $h$ and the remaining one is $l$, i.e. $l=1-3h$ ($h = \frac{1-l}{3}$), and (2) three settings are chosen with probability $l$ and the remaining one is $h$, i.e.  $h = 1-3l$.

It is worth noting that the randomness calculated for $I^{MDL}_{\alpha=1}$ found direct application in the state-of-the-art device-independent protocol for randomness amplification of arbitrarily weak seeds against adversaries holding quantum side information \cite{kessler2020device}. 
We illustrate the usefulness of the tilted Hardy test in randomness amplification by showing that an improvement in the generation rate of the DI protocol in \cite{kessler2020device} can be achieved by considering the tilted Hardy tests for $\alpha \neq 1$. Specifically, \textcolor{black}{for $\alpha = \frac{1}{2}$ global randomness of up to $H_{\infty}(A,B|A_0,B_0,E) =2$ bits can be certified as compared to the corresponding value of $\approx 1.23$ bits from the CHSH expression ($\alpha=1$) used in \cite{acin2012randomness,kessler2020device}. However, the robustness performance for $\alpha = \frac{1}{2}$ is not very good as seen in Fig.~\ref{mdl-l}. The robustness property for $\alpha = \frac{4}{5}$ is comparable with $\alpha = 1$, and from Fig.~\ref{mdl-l}} we see that the certified min-entropy at $I_{\alpha = \frac{4}{5}}^{MDL}$ exceeds that derived for $I_{\alpha = 1}^{MDL}$ for the whole range of $l \in [0, \frac{1}{4}]$ in either case. Therefore, plugging this min-entropy per round into the state-of-the-art DI protocol against quantum adversaries in \cite{kessler2020device} leads to a concomitant improvement in the generation rate of that protocol.

\subsection{The guessing probability of a classical adversary allowed to prepare arbitrary no-signaling boxes for Alice and Bob}\label{nosig}

In this section, we compute the upper bound of the guessing probability of a classical adversary allowed to prepare arbitrary no-signaling boxes $P_{A,B|X,Y}$ in the scenario where Alice and Bob observe some (noisy) value $I^{MDL}_{\alpha}(P_{A,B|X,Y}) = \delta$. This scenario has direct applicability to the fully general scenario of arbitrary no-signaling adversaries, i.e., in the scenario of arbitrary tripartite no-signaling boxes $P_{A,B,E|X,Y,Z}$ with a concomitant loss in the generation rate as shown in \cite{brandao2016realistic}.

\color{black}

\begin{proposition} Suppose that the honest parties observe a value $I_{\alpha}^{MDL}(P_{A,B|X,Y}) = \delta$ for the MDL quantity defined in \eqref{MDL} \textcolor{black}{using a SV-source with $l\leq P_{X,Y}(x,y|\lambda)\leq h,\forall x,y\in\{0,1\}$ }. Then for any pair of inputs $x,y$ and outcomes $a,b$, their shared box satisfies
 \begin{equation}
  \begin{split}
    P_{A,B|X,Y}(a,b|A_x, B_y) &\leq 1-\frac{\delta}{lh\alpha}\qquad \alpha\in(\frac{1}{3},1]\\
    P_{A,B|X,Y}(a,b|A_x, B_y) &\leq 1-\frac{\delta}{lh} \qquad  \alpha\in(1,+\infty)
  \end{split}
\end{equation}
\end{proposition}
The proof of this proposition can be found in Appendix~\ref{bound_prob}. As mentioned earlier, the above bound finds direct applicability in the state-of-the-art protocol for randomness amplification of arbitrary $\epsilon$-SV sources against general no-signaling adversaries. An optimization over $\alpha$ and the amount of shared entanglement is then required to identify the best protocol for randomness amplification against no-signaling adversaries in terms of these two parameters.











\section{Achieving Global randomness of $2\log d$ bits with arbitrarily limited measurement independence}\label{gadgetsec}
In this section, we present candidate Hardy paradoxes that potentially allow for the certification of the maximal amount of $\textcolor{black}{H^{\text{global}}_{\infty}}=2 \log d$ bits of global randomness for quantum systems of local Hilbert space dimensions $2, 4$ or $8$. The candidate Hardy test for systems of local dimension $2$ is the well-known ladder Hardy test \cite{boschi1997ladder} while the candidate Hardy tests for systems of dimensions $4, 8$ are novel constructions based on a connection to sub-structures of Kochen-Specker tests in \cite{ramanathan2020gadget}.  

We first recall the Ladder version of the Hardy test. Here, Alice and Bob perform measurements corresponding to $N+1$ inputs $x,y\in\{0,1\ldots,N\}$ and obtain binary outputs  $a,b\in\{0,1\}$ respectively. The Ladder Hardy test is comprised of $2N+1$ Hardy zero constraints: 
\begin{equation}\label{Hardy_paradox_ladder}
  \begin{split}
   P_{A,B|X,Y}(0,1|A_k,B_{k-1})&=0\qquad \forall k \in\{1,\ldots,N \}\\
   P_{A,B|X,Y}(1,0|A_{k-1},B_k)&=0\qquad \forall k \in\{1,\ldots,N \}\\
   P_{A,B|X,Y}(0,0|A_0,B_0)&=0 \\
  \end{split}
\end{equation}
and the (non-zero) Hardy probability is $P_{Hardy}=P_{A,B|X,Y}(0,0|A_N,B_N)$. In classical (LHV) theories, the $2N+1$ Hardy constraints imply that the Hardy probability is zero, i.e., $P_{Hardy}=0$. However, quantum theory allows to reach $P_{Hardy}\to\frac{1}{2}$ under the following optimal quantum strategy~\cite{ramanathan2021no}. 
Alice and Bob share the two-qubit state $|\psi\rangle=\alpha|00\rangle-\beta|11\rangle$, where $\alpha,\beta$ are parameters depending on on $N$ with $|\alpha|^2+|\beta|^2=1$, we denote their ratio as $t=\frac{\alpha}{\beta}$. The projectors for the two parties correspond to $\{|A_k^0 \rangle= \cos a_k |0\rangle +\sin a_k |1\rangle, |A_k^1 \rangle= -\sin a_k |0\rangle +\cos a_k |1\rangle\}$ and $\{|B_k^0 \rangle= \cos b_k |0\rangle +\sin b_k |1\rangle,
|B_k^1 \rangle= -\sin b_k |0\rangle +\cos b_k |1\rangle\}$ respectively, where $a_k, b_k \in [0,2\pi)$ are the angles for Alice and Bob's $k$-th measurement.
For the state satisfying all the Hardy constraints and achieving the optimal Hardy probability,  one can calculate the probability distributions for any measurement setting $x,y$ as
\begin{equation}\label{prob_dis}
  \begin{split}
  P_{A,B|X,Y}(0,0|A_x,B_y)&=\frac{t^2}{1+t^2}\frac{(1-(-1)^{2N-x-y}\cdot t^{x+y})^2}{(1+t^{2x+1})(1+t^{2y+1})}\\
  P_{A,B|X,Y}(0,1|A_x,B_y)&=\frac{t^2}{1+t^2}\frac{(1+(-1)^{y-x}\cdot t^{x-y-1})^2\cdot t^{2y+1}}{(1+t^{2x+1})(1+t^{2y+1})}\\
P_{A,B|X,Y}(1,0|A_x,B_y)&=\frac{t^2}{1+t^2}\frac{(1+(-1)^{x-y}\cdot t^{y-x-1})^2\cdot t^{2x+1}}{(1+t^{2x+1})(1+t^{2y+1})}\\
P_{A,B|X,Y}(1,1|A_x,B_y)&=\frac{1}{1+t^2}\frac{(1-(-1)^{2N-x-y}\cdot t^{x+y+2})^2}{(1+t^{2x+1})(1+t^{2y+1})}
\end{split}
\end{equation}
The Hardy probability is then given by
\begin{equation}\label{hardy_violation_appe}
\begin{split}
  P_{Hardy}=&P_{A,B|X,Y}(0,0|A_N,B_N) \\
  =&\max_{0\leq t\leq 1}\frac{t^2}{1+t^{2}}\left(\frac{1-t^{2N}}{1+t^{2N+1}} \right)\to \frac{1}{2}\quad \textcolor{black}{(N\to\infty,t\to 1_{-})}
  \end{split}
\end{equation}
Under the (as yet unproven) assumption that this optimal quantum strategy \textcolor{black}{($N\to\infty,t\to 1_{-}$)} is self-testing, we see that one may certify arbitrarily close to the maximum possible value of $2$ bits of global randomness from the measurement setting $x=N,y=0$ (or $x=0,y=N$), i.e., $P_{A,B|X,Y}(a,b|A_N,B_0) \to \frac{1}{4}$ for all $a, b \in \{0,1\}$ \textcolor{black}{(note that both the two limits $N\to\infty,t\to 1_{-}$ need to be taken.)}

The second family of Hardy tests (for certifying $2 \log_2 d$ bits of global randomness in $d=4,8$) is based on a novel construction using sub-structures of Kochen-Specker sets, that were termed $01$-gadgets in \cite{ramanathan2020gadget}. A $01$-gadget is a graph $G = (V_G, E_G)$ with a faithful orthogonal representation in $\mathbb{C}^{\omega(G)}$, where $\omega(G)$ denotes the clique number of the graph. Here, a faithful orthogonal representation is a mapping of the vertices of the graph to (unit) vectors such that adjacent vertices are mapped to orthogonal vectors, and distinct vertices are assigned distinct vectors. Kochen-Specker sets concern $\{0,1\}$-colorings of the graph (i.e., mappings $f: V_G \rightarrow \{0,1\}$) that satisfy the conditions that (i) any two adjacent vertices cannot be assigned value $1$ simultaneously; and that (ii) in any maximal clique, there is a unique vertex that is assigned value $1$. A defining property of a $01$-gadget $G$ is that there exist two distinct \textit{non-adjacent} vertices $v_1 \nsim v_2$ that cannot be assigned value $1$ simultaneously in any feasible  $\{0,1\}$-coloring of $G$, while at the same time, a faithful orthogonal representation of $G$ exists in $\mathbb{C}^{\omega(G)}$ in which $v_1$ and $v_2$ are assigned non-orthogonal vectors.


\begin{figure}[htbp]
\centering
\subfigure[]{
\begin{minipage}[t]{0.5\linewidth}
\centering
\includegraphics[width=0.8\textwidth]{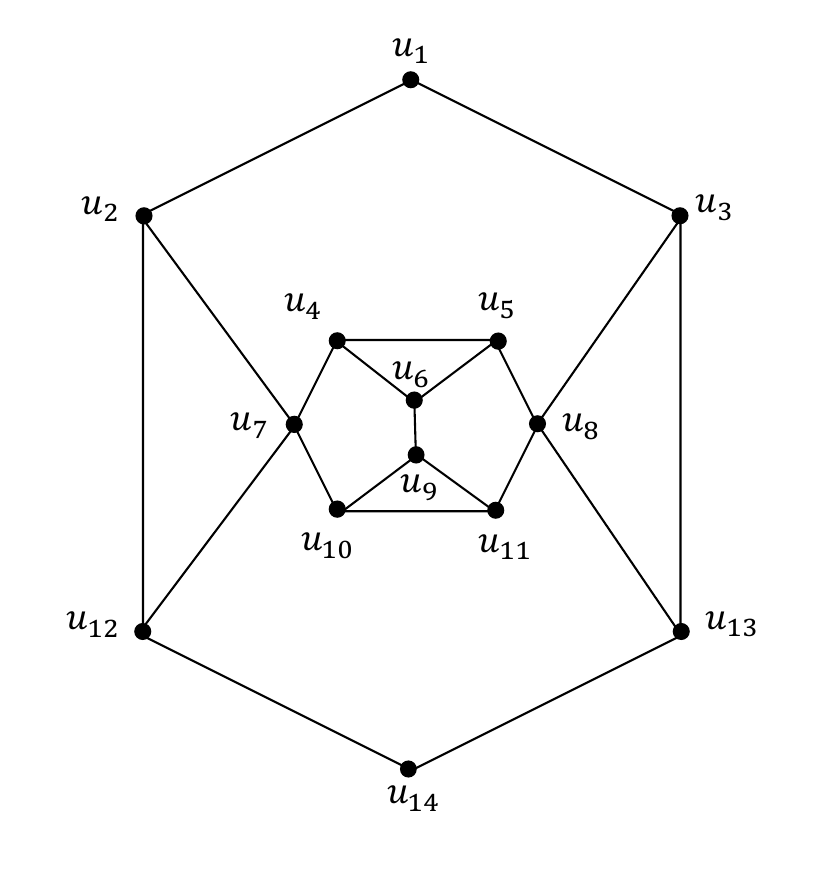}
\label{gadget1}
\end{minipage}%
}%
\subfigure[]{
\begin{minipage}[t]{0.5\linewidth}
\centering
\includegraphics[width=0.85\textwidth]{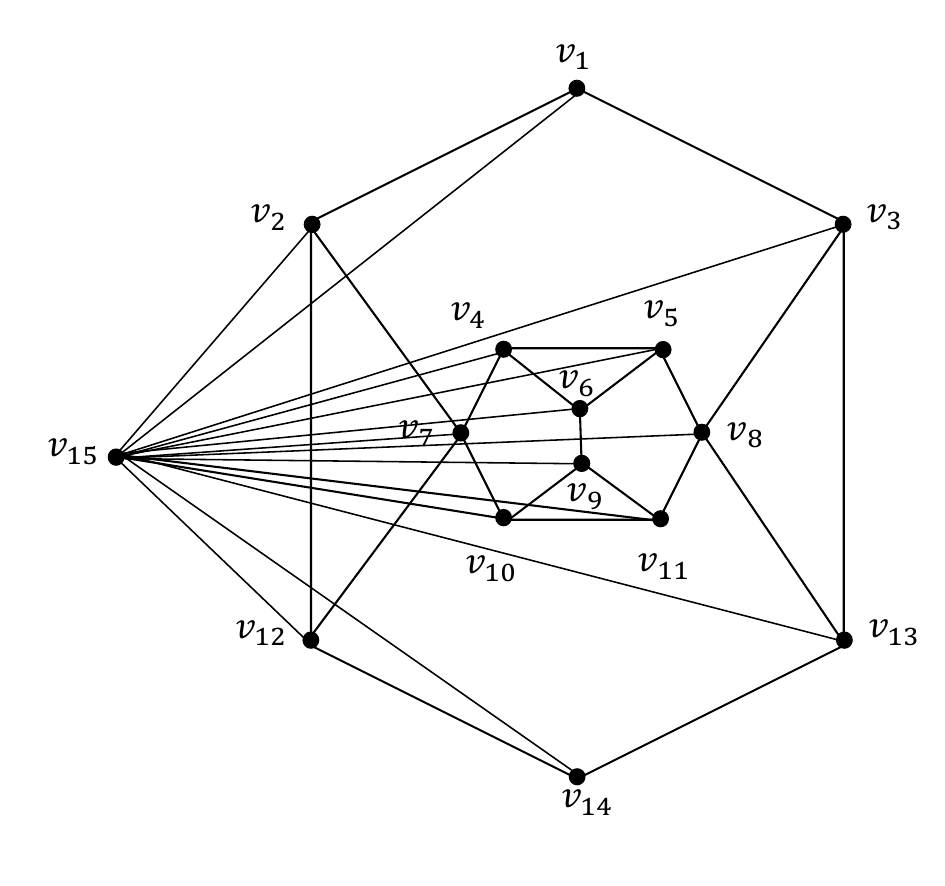}
\label{gadget2}
\end{minipage}%
}
\centering
\caption{ The graph in (a) is a 01-gadget that has a faithful orthogonal representation in $\mathbb{R}^{3}$, and the overlap of the two distinguished vectors $|\langle u_1|u_{14}\rangle|\leq\frac{1}{2}$. The graph in (b) is a $01$-gadget constructed by adding an additional vector $|v_{15}\rangle$ that is orthogonal with all of the vectors in the faithful representation of (a). The graph in (b) has an orthogonal representation in $\mathbb{R}^{4}$, and the maximal overlap of the two distinguished vectors is still $\frac{1}{2}$, i.e., $|\langle v_1|v_{14}\rangle|\leq\frac{1}{2}$.  }
\label{gadget}
\end{figure}

\begin{figure}[htbp]
    \centering
    \includegraphics[width=0.75\textwidth]{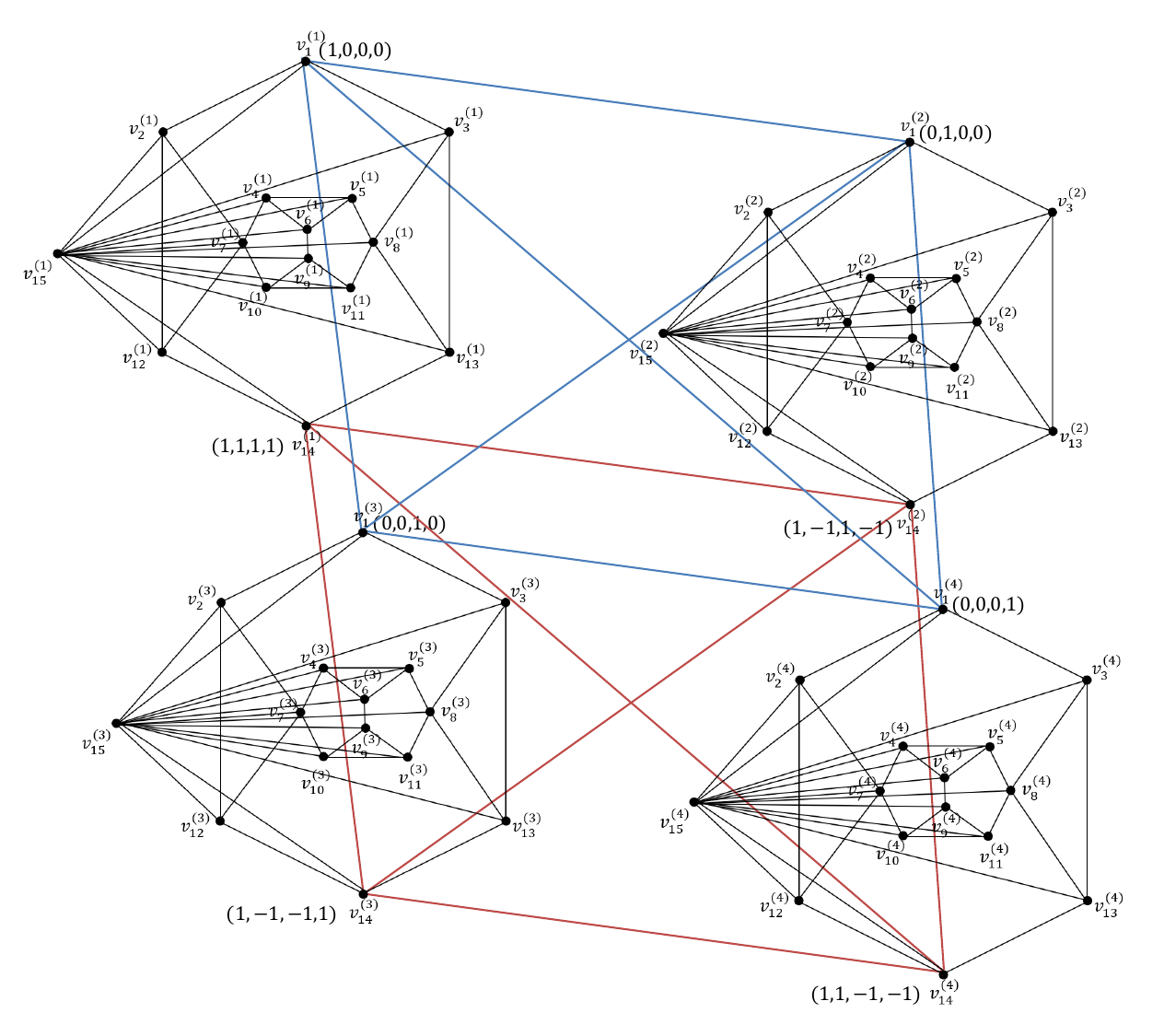}
    \caption{As explained in the text, one can rotate the vectors in the 01-gadget Fig.~\ref{gadget2} to create four copies of it. The corresponding vectors in each copy are mutually orthogonal, i.e., $|v_{k}^{(1)}\rangle,|v_{k}^{(2)}\rangle,|v_{k}^{(3)}\rangle,|v_{k}^{(4)}\rangle$ form a complete basis (the vertices form a maximal clique).  }
    \label{hardy_4bits}
\end{figure}

The figure shown in Fig.~\ref{gadget1} is an example of 01-gadget, which has an orthogonal representation in $\mathbb{R}^{3}$ , and the maximal overlap of the two distinguished vectors  $|u_1\rangle,|u_{14}\rangle$ is $\frac{1}{2}$, i.e., $|\langle u_1|u_{14}\rangle|\leq\frac{1}{2}$ \cite{ramanathan2020gadget}. 
The 01-gadget shown in Fig.~\ref{gadget2} has an orthogonal representation in $\mathbb{R}^{4}$ which is constructed by adding an extra vector $|v_{15}\rangle$ to the first 01-gadget and making this vector  $|v_{15}\rangle$ orthogonal to all of the other vectors. 
 Specifically, since the first 01-gadget has an orthogonal representation in $\mathbb{R}^{3}$, one can keep the structure of the first 01-gadget in the first three dimensions of $\mathbb{R}^{4}$ and the extra vector is the basis vector on the fourth dimension $(0,0,0,1)^{T}$. It is easy to verify that the two distinguished vectors of this new 01-gadget are still  $|v_1\rangle,|v_{14}\rangle$ and their maximal overlap is still $\frac{1}{2}$. Specifically, an orthogonal representation of the 01-gadget Fig.~\ref{gadget2} for which the maximal overlap of the  two special distinguished vertices is $|\langle v_1|v_{14}\rangle|=\frac{1}{2}$ is given as (excluding normalization factors):
\begin{equation}\label{realization}
\begin{aligned}
&\langle v_1 | = (1, 0,0,0),\ 
\langle v_2 | = (0,4,1,1),\  
\langle v_3 | =(0,0,1,1), \\ 
&\langle v_4 |= (-3-\sqrt{3},-5+\sqrt{3},1+\sqrt{3},1+\sqrt{3}), \ 
\langle v_5 | = (3+\sqrt{3}, -3-\sqrt{3},3-\sqrt{3},3-\sqrt{3}), \\ 
&\langle v_6| = (3-\sqrt{3},1+\sqrt{3},1+\sqrt{3},1+\sqrt{3}), \
\langle v_7 | = (-3,1,-2,-2),\
\langle v_8|= (1, 1, 0,0), \\
&\langle v_9 |= (3+\sqrt{3},1-\sqrt{3},1-\sqrt{3},1-\sqrt{3}), \
\langle v_{10} |= (-3+\sqrt{3},-5-\sqrt{3},1-\sqrt{3},1-\sqrt{3}), \\
&\langle v_{11} |= (-3+\sqrt{3},3-\sqrt{3},-3-\sqrt{3},-3-\sqrt{3}), \\
&\langle v_{12} |= (3,1,-2,-2), \
\langle v_{13}|= (-1,1,0,0), \
\langle v_{14} |= (1,1,1,1), \
\langle v_{15} |= (0,0,1,-1).\\
\end{aligned}
\end{equation}
In order to certify 4 bits of global randomness for the quantum system of dimension $4$, we create four copies of this 01-gadget  $G^{(1)}, G^{(2)}, G^{(3)}, G^{(4)}$ in $\mathbb{C}^{4}$ such that the corresponding vectors in each copy are mutually orthogonal, i.e., $|v_{k}^{(1)}\rangle,|v_{k}^{(2)}\rangle,|v_{k}^{(3)}\rangle,|v_{k}^{(4)}\rangle$ form a complete basis (the vertices form a maximal clique). This is achieved by rotating each vector $(a, b, c, d) \in \mathbb{R}^{4}$ in the original 01-gadget Fig.~\ref{gadget2} into the orthogonal vectors $(b,-a,-d, c),(c, d,-a,-b),(d,-c, b,-a)$ in the copies while preserving the orthogonality structure. Note that such a construction is also possible in dimension $8$, because there exist division algebras (the quaternions and octonions) in dimensions $4$ and $8$ \cite{cohn2012basic}. Let $G$ in Fig.~\ref{hardy_4bits} denote the  orthogonality graph of these four copies with vertices $V(G)=$ $V(G^{(1)}) \cup V(G^{(2)}) \cup V(G^{(3)}) \cup V(G^{(4)})$ and edges formed by the orthogonality constraints. 

 Some of us showed in \cite{ramanathan2022large} that there exists a one-to-one correspondence between $01$-gadget in dimension $d$ and two-player Hardy paradoxes for the maximally entangled state in $\mathbb{C}^{d} \otimes \mathbb{C}^{d}$. We now proceed to construct the Hardy paradox for the maximally entangled state $|\psi \rangle = \frac{1}{2} \sum_{i=0}^{3} | i, i \rangle$. 
 To do this, we first complete all the possible measurement bases (maximal cliques in the graph), these bases will serve as the measurement inputs $x,y$ for both players. In particular, let us denote by $x^*,y^*$ the respective maximal cliques formed by the distinguished vertices $v_1^{(1)},v_1^{(2)},v_1^{(3)},v_1^{(4)}$ (the blue clique in Fig.~\ref{hardy_4bits}) and $v_{14}^{(1)},v_{14}^{(2)},v_{14}^{(3)},v_{14}^{(4)}$ (the red clique in Fig.~\ref{hardy_4bits}). The outcomes of each measurement are labeled by  $a,b\in\{1,2,3,4\}$ for the two players respectively, and correspond to the vertices in the clique. For convenience, let us rewrite the notation of the orthogonal representations of $G$ as  $\{|v_{z,c}\rangle\}\quad z\in\{x,y\},c\in\{a,b\}$.
 We define the set $S_{Hz}$ of Hardy `zero' constraints in this case as
\begin{equation}\label{hardy_zero}
S_{Hz}:=\{((x, a),(y, b)):\langle v_{(x, a)} |v_{(y, b)}\rangle=0\} \qquad \forall x,y,a,b
\end{equation}
The Hardy probability is then given by the sum of the probabilities of the input-output pairs in the set $S_{exp}$ defined as 
    $S_{exp} := \cup_{k=1}^{4}\{((x^{*}, v_{1}^{(k)}),(y^{*}, v_{14}^{(k)}))\}$. 
Note that these input-output pairs correspond to the four pairs of distinguished vertices of the four copies of the $01$-gadget. It is readily seen that in any deterministic classical (LHV) behavior, by the property of the $01$-gadget that the distinguished vertices cannot be assigned value $1$ simultaneously, it holds that the sum of the probabilities of the input-output pairs in $S_{exp}$ is always $0$.  


On the other hand, we present a quantum strategy that satisfies all the zero constraints while simultaneously achieving the Hardy probability of $1$. In the strategy, the two players share a two-ququart maximally entangled state $|\psi \rangle= \frac{1}{2} \sum_{i=0}^{3}|i, i\rangle$ and upon receiving the input $x,y$ they measure in the corresponding bases given by the orthogonal representation: $\Pi_{x}=\{|v_{(x, a)}\rangle\langle v_{(x, a)}|\}_{a=1, \ldots, 4}$  and $\Sigma_{y}=\{|v_{(y, b)}\rangle\langle v_{(y, b)}|\}_{b=1, \ldots, 4}$  and return the outcomes $a,b \in \{1,\dots,4\}$. We see that for all $((x, a),(y, b)) \in S_{Hz}$:
\begin{equation}\label{hardy_condition}
\langle\psi |(|v_{(x, a)}\rangle \otimes|v_{(y, b)}\rangle)=\frac{1}{2} \overline{\langle v_{(x, a)} \mid \bar{v}_{(y, b)}\rangle}=0 
\end{equation}
which correspond to the zero constraints in the Hardy paradox. While, for the input-output pairs in $S_{exp}$, we obtain
\begin{equation}\label{hardyprobability}
P_{A,B|X,Y}\left(v_{1}^{(k)}, v_{14}^{(k)} | x^{*}, y^{*}\right)=\frac{1}{4}|\langle v_{1}^{(k)} |\bar{v}_{14}^{(k)}\rangle|^{2}=\frac{1}{16} \qquad \text{for} \; k=1,\ldots,4
\end{equation}
which correspond to the non-zero Hardy expressions in the  Hardy paradox. Furthermore, from the orthogonal representation shown in \eqref{realization}, we also see that the outcomes of the measurement settings $x^*,y^*$ are uniform:
\begin{equation*}
P_{A,B|X,Y}\left(v_{1}^{(i)}, v_{14}^{(j)} | x^{*}, y^{*} \right)=\frac{1}{4}|\langle v_{1}^{(i)} |\bar{v}_{14}^{(j)}\rangle|^{2}=\frac{1}{16}\quad \forall i,j=1,\ldots, 4
\end{equation*}
Thus, the presented Hardy test is a good candidate to certify the maximum possible value (using systems of local Hilbert space dimension $d = 4$) of $4$ bits of global randomness and $2$ bits of local randomness from the measurement settings $x^*,y^*$. We defer the detailed self-testing statement of the above Hardy test and the corresponding test using systems of local dimension $d=8$ to forthcoming work.




\section{Conclusions}
The generalisation of device-independent protocols for randomness generation, key distribution and self-testing to the scenario when the honest parties do not have trusted, fully uniform private seeds remains an interesting open question. In this paper, we have proposed a class of tilted Hardy tests in the simplest Bell scenario of two players performing a pair of binary measurements on two-qubit states that serve as ideal candidates for this generalisation. Our derived results on the amount of certifiable randomness from these tests against different sets of adversaries find immediate application in the state-of-the-art protocols for randomness amplification of weak SV seeds against quantum and no-signaling adversaries. While the tests allow to certify the maximum possible amount of $1$ bit of local randomness for each player, they fall short of the ideal global randomness of $2$ bits from the measurement outputs of both players. To remedy this defect, we proposed a class of Hardy tests in systems of local Hilbert space dimensions $4, 8$ that can allow to certify up to the maximum possible $2 \log_2 d$ bits of randomness under arbitrarily limited measurement independence. It is an interesting task for future research to utilize the proposed tests to improve the generation rates of fully device-independent protocols against general adversaries.      
\section{Acknowledgments}
We acknowledge support from the Start-up Fund ”Device-Independent Quantum Communication Networks” from The University of Hong Kong, the Seed Fund ”Security of Relativistic Quantum Cryptography” (Grant No. 201909185030),  the Early Career Scheme (ECS) grant ”Device-Independent Random Number Generation and Quantum Key Distribution with Weak Random Seeds” (Grant No. 27210620), the General Research Fund (GRF) grant "Semi-device-independent cryptographic applications of a single trusted quantum system" (Grant No. 17211122) and the Research Impact Fund (RIF) "Trustworthy quantum gadgets for secure online communication" (Grant No. R7035-21). P.H. acknowledges support from the Foundation for Polish Science (IRAP project, ICTQT, contract no. 2018/MAB/5, co-financed by EU within the Smart Growth Operational Programme).


\begin{appendix}
\section{Self-testing general pure two-qubit entangled states with the tilted Hardy test}\label{proof-w-hardy}
In this section, we provide the proof of Prop.~\ref{prop:w-hardy}. For convenience, we review this Proposition here. 
\begin{proposition*}
Suppose that the Hardy zero constraints  $P_{A,B|X,Y}(0,1|A_0,B_1)=\linebreak P_{A,B|X,Y}(1,0|A_1,B_0)$ $=P_{A,B|X,Y}(0,0|A_1,B_1)=0$ are satisfied. Then in any local hidden variable theory it holds that $P_{Hardy}^{w,LHV}\leq\max(0,w)$. On the other hand, quantum theory allows to achieve $P_{Hardy}^{w,Q}=\frac{(4w+5)\sqrt{4w+5}-(12w+11)}{2(w+1)}$. Furthermore, up to local isometries, the value is uniquely achieved by the two-qubit state $|\psi_{\theta}\rangle=\cos(\frac{\theta}{2})|00\rangle-\sin(\frac{\theta}{2})|11\rangle$ with $\theta=\arcsin{(3-\sqrt{4w+5})}$ and the measurement $\{A_0,A_1\}$, $\{B_0,B_1\}$ given by 
\begin{equation}\label{eq:w-measurements}
  \begin{split}
   A_0=B_0&=\frac{-(2+\sin\theta)(1-\sin\theta)^{\frac{1}{2}}}{(2-\sin\theta)(1+\sin\theta)^{\frac{1}{2}}}\cdot Z+\frac{-\sqrt2\sin\theta(\sin\theta)^{\frac{1}{2}}}{(2-\sin\theta)(1+\sin\theta)^{\frac{1}{2}}}\cdot X,\\
  A_1=B_1&=\frac{-(1-\sin\theta)^{\frac{1}{2}}}{(1+\sin\theta)^{\frac{1}{2}}}\cdot Z+\frac{\sqrt2(\sin\theta)^{\frac{1}{2}}}{(1+\sin\theta)^{\frac{1}{2}}}\cdot X, 
\end{split}
\end{equation}
where $X$ and $Z$ are Pauli matrices.
\end{proposition*}
\begin{proof}
The proof consists of the following three statements.

\begin{itemize}

    \item The classical value $P_{Hardy}^{w,LHV}\leq\max(0,w)$.
    
        Any local hidden variable behavior can be written as a convex combination of deterministic non-signaling behaviors $\{P_{A,B|X,Y}(a,b|x,y)\}$. In the Bell scenario under consideration, it is well known that there are $16$ deterministic behaviors in total (where each party deterministically outputs a specific value of $a, b \in \{0,1\}$ for each input). Of these, there are five deterministic non-signaling behaviors that meet the Hardy `zero' constraints $P_{A,B|X,Y}(0,1|A_0,B_1)=P_{A,B|X,Y}(1,0|A_1,B_0)=P_{A,B|X,Y}(0,0|A_1,B_1)=0$, shown in Table~\ref{table_hardy1} and \ref{table_hardy2}. As is easily seen, the three Hardy zero constraints force $P_{A,B|X,Y}(0,0|A_0,B_0) = 0$ in each of the deterministic non-signaling boxes. $P_{A,B|X,Y}(1,1|A_0,B_0) \leq 1$ leads to the bound of the non-zero Hardy probability term: $P_{Hardy}^{w,LHV} = P_{A,B|X,Y}(0,0|A_0,B_0) + w P_{A,B|X,Y}(1,1|A_0,B_0) \leq\max(0,w)$. 
        
        
\begin{table}[ht]\centering
\caption{Local boxes that lead to the upper bound of tilted Hardy's paradox when $w\geq 0$. }\label{table_hardy1}
\begin{tabular}{cccccccccc}
  \begin{tabular}{cc|cc|cc|}
  \multicolumn{2}{c}{~} &\multicolumn{2}{c}{$B_0$}& \multicolumn{2}{c}{$B_1$}\\
  \multicolumn{2}{c}{~} &\multicolumn{2}{c}{$\overbrace{0\quad 1}$}&\multicolumn{2}{c}{$\overbrace{0\quad 1}$} \\
  \cline{3-6}
  &\multirow{2}{1cm}{$A_0\left\{
                      \begin{array}{l}
                        0  \\
                        1 
                      \end{array}
                    \right.$}&0 & 0 & 0&0 \\
  & &0 & 1 & 0 &1\\
  \cline{3-6}
  & \multirow{2}{1cm}{$A_1\left\{
                      \begin{array}{l}
                        0  \\
                        1
                      \end{array}
                    \right.$}&0 & 0 & 0&0 \\
  & &0 & 1 & 0 &1\\
  \cline{3-6}
\end{tabular} & & & \begin{tabular}{c|cc|cc|}
  \multicolumn{1}{c}{}&\multicolumn{2}{c}{$B_0$}& \multicolumn{2}{c}{$B_1$}\\
  \multicolumn{1}{c}{}&\multicolumn{2}{c}{$\overbrace{0\quad 1}$}&\multicolumn{2}{c}{$\overbrace{0\quad 1}$} \\
  \cline{2-5}
  &0 & 0 & 0&0 \\
  &0 & 1 & 1 &0\\
  \cline{2-5}
  \multirow{2}{0cm}{}&0 & 0 & 0&0 \\
  &0 & 1 & 1 &0\\
  \cline{2-5}
\end{tabular} & & & \begin{tabular}{c|cc|cc|}
  \multicolumn{1}{c}{}&\multicolumn{2}{c}{$B_0$}& \multicolumn{2}{c}{$B_1$}\\
  \multicolumn{1}{c}{}&\multicolumn{2}{c}{$\overbrace{0\quad 1}$}&\multicolumn{2}{c}{$\overbrace{0\quad 1}$} \\
  \cline{2-5}
  &0 & 0 & 0&0 \\
  &0 & 1 & 0 &1\\
  \cline{2-5}
  &0 & 1 & 0&1 \\
  &0 & 0 & 0 &0\\
  \cline{2-5}
\end{tabular} \\
 & & &  \\
\end{tabular}
\end{table}

\begin{table}[ht]\centering
\caption{Local boxes that lead to the upper bound of tilted Hardy's paradox when $w\leq 0$.}\label{table_hardy2}
\begin{tabular}{cccccccc}
  \begin{tabular}{cc|cc|cc|}
  \multicolumn{2}{c}{~} &\multicolumn{2}{c}{$B_0$}& \multicolumn{2}{c}{$B_1$}\\
  \multicolumn{2}{c}{~} &\multicolumn{2}{c}{$\overbrace{0\quad 1}$}&\multicolumn{2}{c}{$\overbrace{0\quad 1}$} \\
  \cline{3-6}
  &\multirow{2}{1cm}{$A_0\left\{
                      \begin{array}{l}
                        0  \\
                        1
                      \end{array}
                    \right.$}&0 & 1 & 1&0 \\
  & &0 & 0 & 0 &0\\
  \cline{3-6}
  & \multirow{2}{1cm}{$A_1\left\{
                      \begin{array}{l}
                        0  \\
                        1
                      \end{array}
                    \right.$}&0 & 0 & 0&0 \\
  & &0 & 1 & 1 &0\\
  \cline{3-6}
\end{tabular} & & & 
\begin{tabular}{c|cc|cc|}
  \multicolumn{1}{c}{}&\multicolumn{2}{c}{$B_0$}& \multicolumn{2}{c}{$B_1$}\\
  \multicolumn{1}{c}{}&\multicolumn{2}{c}{$\overbrace{0\quad 1}$}&\multicolumn{2}{c}{$\overbrace{0\quad 1}$} \\
  \cline{2-5}
  &0 & 0 & 0&0 \\
  &1 & 0 & 0 &1\\
  \cline{2-5}
  \multirow{2}{0cm}{}&1 & 0 & 0&1 \\
  &0 & 0 & 0 &0\\
  \cline{2-5}
\end{tabular}
\end{tabular}
\end{table}
    
    \item The quantum value $P_{Hardy}^{w,Q}=\frac{(4w+5)\sqrt{4w+5}-(12w+11)}{2(w+1)}$.

        In the Bell scenario under consideration where each party performs measurements of two binary observables, the well-known Jordan's Lemma \cite{jordan1875essai} states that arbitrary Hermitian observables can be block diagonalized into blocks of size at most $2 \times 2$.  That is, for Alice and Bob's measurement observables $A_x,B_y$ and $x,y\in\{0, 1\}$ , the corresponding projectors can be decomposed as  $A_{a|x}=\bigoplus_i A_{a|x}^i$ and $B_{b|y}=\bigoplus_j B_{b|y}^j$, where $A_{a|x}^i,B_{b|y}^j$ are the corresponding projectors acting on Hilbert spaces $\mathbf{H}_A^i$ and $\mathbf{H}_B^j$ of dimensions $d \leq 2$ for all $a, b$ and $x, y$. The joint probability of outcomes $a, b$ given measurements $x, y$ performed on the quantum system can be decomposed as
        \begin{equation}
        \begin{split}
        P_{A,B|X,Y}(a,b|A_x,B_y)&=\text{Tr}[(A_{a|x}\otimes B_{b|y})\rho] \\
        &=\sum_{i,j} q_{i,j} \text{Tr}[(A_{a|x}^i\otimes B_{b|y}^j)\rho_{i,j}]\\
        &= \sum_{i,j} q_{i,j} P^{i,j}_{A,B|X,Y}(a,b|A_x,B_y),
        \end{split}
        \end{equation}
        where $\rho_{i,j}$ is a (two-qubit) projection of the shared quantum state $\rho$ on Hilbert spaces $\mathbf{H}_A^i\otimes \mathbf{H}_B^j$,
        $q_{i,j}$ is the corresponding coefficient with $\sum_{i,j} q_{i,j}=1$.
        When the three Hardy zero constraints $P_{A,B|X,Y}(0,1|A_0,B_1)=P_{A,B|X,Y}(1,0|A_1,B_0)=P_{A,B|X,Y}(0,0|A_1,B_1)=0$ are satisfied, following \cite{hardy1993nonlocality}, it is easy to calculate the quantum state and measurements which achieve the maximal value of the Hardy probability $P_{Hardy}^{w,Q}$. For an arbitrary partially entangled state $|\psi_{\theta}\rangle=\cos(\frac{\theta}{2})|00\rangle-\sin(\frac{\theta}{2})|11\rangle$ with $\theta \in (0,\frac{\pi}{2})$, $P_{Hardy}^{w,Q}$ is given by:
        \begin{align}\label{hardy_value_vio}
        &P_{Hardy}^{w,Q}=P_{A,B|X,Y}(0,0|A_0,B_0)+w\cdot P_{A,B|X,Y}(1,1|A_0B_0) \nonumber \\
        &=\left[\frac{(\cos(\frac{\theta}{2})\sin(\frac{\theta}{2}))(\cos(\frac{\theta}{2})-\sin(\frac{\theta}{2}))}{1-\cos(\frac{\theta}{2})\sin(\frac{\theta}{2})}\right]^2
        +w \left[\frac{1-2\cos(\frac{\theta}{2})\sin(\frac{\theta}{2})}{(\cos(\frac{\theta}{2})-\sin(\frac{\theta}{2}))(1-\cos(\frac{\theta}{2})\sin(\frac{\theta}{2}))}\right]^2 \nonumber \\
        &=\frac{(\sin^2\theta+4\cdot w)(1-\sin\theta)}{(2-\sin\theta)^2}.
        \end{align}

        Setting $\frac{d P_{Hardy}^{w,Q}}{d \sin\theta} = 0$ gives $\sin\theta =3-\sqrt{4w+5}$. It is then readily checked that this point gives the maximum of the Hardy probability as:  
        \begin{equation}
        P_{Hardy}^{w,Q}= \frac{(4w+5)\sqrt{4w+5}-(12w+11)}{2(w+1)}
        \end{equation}
        for any $w\in(-\frac{1}{4},1)$. 
        For each pure two-qubit entangled state $|\psi_{\theta} \rangle$ other than the singlet (with $\theta = \pi/2$), we see that there exists a corresponding $w$ (and thus, a corresponding tilted Hardy test) such that $| \psi_{\theta} \rangle$ gives the optimal violation of that test.

    \item Self-testing of $|\psi_{\theta} \rangle$ and $\{A_x\}$, $\{B_y\}$.

        Without loss of generality, choosing the  computational basis $\{|2i\rangle\langle2i|,~|2i+1\rangle\langle2i+1|\}$ on Hilbert space $\mathbf{H}_A^i$  for Alice and the  computational basis $\{|2j\rangle\langle2j|,~|2j+1\rangle\langle2j+1|\}$ on Hilbert space $\mathbf{H}_B^j$ for Bob, the shared quantum state can be expressed as \textcolor{black}{$\rho_{i,j,\theta}=|\psi_{i,j,\theta}\rangle\langle\psi_{i,j,\theta}|$,
        where
        $|\psi_{i,j,\theta}\rangle=\cos (\frac{\theta}{2})|2i,2j\rangle-\sin (\frac{\theta}{2})|2i+1,2j+1\rangle$.
        According to the above discussion, the Hardy zero constraints are satisfied and the  upper bound of the Hardy expression $P_{Hardy}^{w,Q}$ is attained if and only if $\sin\theta=3-\sqrt{4w+5}$ and the operators are:
        \begin{align} 
          {A_{1}^i}&=c_{1z}\cdot(|2i\rangle\langle2i|-|2i+1\rangle\langle2i+1|)
                +c_{1x}\cdot(|2i\rangle\langle2i+1|+|2i+1\rangle\langle2i|) \nonumber \\
          {B_{1}^j}&=c_{1z}\cdot(|2j\rangle\langle2j|-|2j+1\rangle\langle2j+1|)
                +c_{1x}\cdot(|2j\rangle\langle2j+1|+|2j+1\rangle\langle2j|)\nonumber \\
          {A_{0}^i}&=c_{0z}\cdot(|2i\rangle\langle2i|-|2i+1\rangle\langle2i+1|)
                +c_{0x}\cdot(|2i\rangle\langle2i+1|+|2i+1\rangle\langle2i|) \nonumber \\
          {B_{0}^j}&=c_{0z}\cdot(|2j\rangle\langle2j|-|2j+1\rangle\langle2j+1|)
                +c_{0x}\cdot(|2j\rangle\langle2j+1|+|2j+1\rangle\langle2j|)
        \end{align}
        with
        \begin{align}
          c_{1z}&=\frac{-(1-\sin\theta)^{\frac{1}{2}}}{(1+\sin\theta)^{\frac{1}{2}}}, \qquad \qquad
          c_{1x}=\frac{\sqrt2(\sin\theta)^{\frac{1}{2}}}{(1+\sin\theta)^{\frac{1}{2}}} \nonumber \\
          c_{0z}&=\frac{-(2+\sin\theta)(1-\sin\theta)^{\frac{1}{2}}}{(2-\sin\theta)(1+\sin\theta)^{\frac{1}{2}}}, \qquad
          c_{0x}=\frac{-\sqrt2\sin\theta(\sin\theta)^{\frac{1}{2}}}{(2-\sin\theta)(1+\sin\theta)^{\frac{1}{2}}}
        \end{align}}
        
        As such, the \textcolor{black}{actual experiment quantum system} can satisfy all the Hardy zero constraints and achieve the bound  $P_{Hardy}^{w,Q}$, if and only if the state is given by
         \textcolor{black}{$ \widetilde{|\psi_\theta\rangle}=\bigoplus_{i, j} \sqrt{q_{i, j}}|\psi_{i, j,\theta}\rangle$
        with the measurements being
          $\widetilde{A_{1}}=\bigoplus_{i}{A_{1}^i}$, $\widetilde{A_{0}}=\bigoplus_{i}{A_{0}^i}$, 
         $\widetilde{B_{1}}=\bigoplus_{j}{B_{1}^j}$ and  $\widetilde{B_{0}}=\bigoplus_{j}{B_{0}^j}$,}
        with coefficients $q_{i,j}$ satisfying $q_{i,j}\geq 0$, $\sum_{i,j} q_{i,j}=1$.
        The isometry is given by $\Phi=\Phi_{A}\otimes \Phi_{B}$ ~\cite{rabelo2012device}  with $\Phi_C(|2k\rangle_C\otimes|0\rangle_{C'})\mapsto |2k\rangle_C\otimes|0\rangle_{C'}$ and 
           $\Phi_C(|2k+1\rangle_{C}\otimes |0\rangle_{C'})\mapsto |2k\rangle_{C}\otimes |1\rangle_{C'}$ for $C = A, B$. 
        With ancillary qubits $|00\rangle_{A'B'}$, \textcolor{black}{ the reference state $|\psi_\theta\rangle$ and reference measurements $A_{1},B_{1},A_{0},B_{0}$ can be self-tested from the physical state $\widetilde{|\psi_\theta\rangle}$ and physical measurements $\widetilde{A_{1}},\widetilde{A_{0}},\widetilde{B_{1}},\widetilde{B_{0}}$}:
        \begin{equation}
          \begin{split}
          &\Phi (\textcolor{black}{\widetilde{|\psi_\theta\rangle}_{AB}}\otimes|00\rangle_{A'B'})\\
          &=\Phi(\bigoplus_{i, j} \sqrt{q_{i j}}(\cos (\frac{\theta}{2})|2i,2j\rangle -\sin (\frac{\theta}{2})|2i+1,2j+1\rangle)_{AB}\otimes|00\rangle_{A'B'})\\
           &=\bigoplus_{i, j} \sqrt{q_{i j}}(\cos (\frac{\theta}{2})|2i,2j\rangle_{AB}\otimes|00\rangle_{A'B'}-\sin(\frac{\theta}{2})|2i,2j\rangle_{AB}\otimes|11\rangle_{A'B'})\\
           &=\bigoplus_{i, j} \sqrt{q_{i j}}|2i,2j\rangle_{AB}\otimes(\cos (\frac{\theta}{2})|00\rangle_{A'B'}-\sin(\frac{\theta}{2})|11\rangle_{A'B'})\\
          &=|\sigma\rangle_{AB}\otimes\textcolor{black}{|\psi_{\theta}\rangle_{A' B'}}
        \end{split}
        \end{equation}
          \begin{equation}
          \begin{split}
          &\Phi(\textcolor{black}{\widetilde{A_{1}}\widetilde{|\psi_{\theta}\rangle}_{AB}}\otimes|00\rangle_{A'B'})\\
          &=\Phi(\bigoplus_{i, j} \sqrt{q_{i j}}[c_{2z}\cdot(|2i\rangle\langle2i|-|2i+1\rangle\langle2i+1|)+c_{2x}\cdot(|2i\rangle\langle2i+1|+|2i+1\rangle\langle2i|)]\\&\qquad(\cos (\frac{\theta}{2})|2i,2j\rangle-\sin (\frac{\theta}{2})|2i+1,2j+1\rangle)_{AB}\otimes|00\rangle_{A'B'})\\
          &=\Phi(\bigoplus_{i, j}\sqrt{q_{i j}} [\cos (\frac{\theta}{2})(c_{2z}|2i,2j\rangle+c_{2x}|2i+1,2j\rangle)\\&\qquad \qquad \quad -\sin (\frac{\theta}{2})(-c_{2z}|2i+1,2j+1\rangle+c_{2x}|2i,2j+1\rangle)]_{AB}\otimes|00\rangle_{A'B'})\\
           &=\bigoplus_{i, j}\sqrt{q_{i j}}[\cos (\frac{\theta}{2})(c_{2z}|2i,2j\rangle_{AB}\otimes|00\rangle_{A'B'}+c_{2x}|2i,2j\rangle\otimes|10\rangle_{A'B'})\\&\qquad\qquad-\sin(\frac{\theta}{2})(-c_{2z}|2i,2j\rangle_{AB}\otimes|11\rangle_{A'B'}+c_{2x}|2i,2j\rangle_{AB}\otimes|01\rangle_{A'B'})]\\
           &=\bigoplus_{i, j}\sqrt{q_{i j}}|2i,2j\rangle_{AB}\otimes [\cos (\frac{\theta}{2})(c_{2z}|00\rangle_{A'B'}+c_{2x}|10\rangle_{A'B'})\\&\qquad\qquad-\sin(\frac{\theta}{2})(-c_{2z}|11\rangle_{A'B'}+c_{2x}|01\rangle_{A'B'})]\\
           &=\bigoplus_{i, j}\sqrt{q_{i j}}|2i,2j\rangle_{AB}\otimes (c_{2z}\cdot Z+c_{2x}\cdot X)(\cos (\frac{\theta}{2})|00\rangle_{A'B'}-\sin(\frac{\theta}{2})|11\rangle_{A'B'})\\
           &=|\sigma\rangle_{AB}\otimes \textcolor{black}{{A_1}|\psi_{\theta}\rangle_{A' B'}}
        \end{split}
        \end{equation}
        and
          \begin{equation}
          \begin{split}
          &\Phi(\textcolor{black}{\widetilde{A_{0}}\widetilde{|\psi_{\theta}\rangle}_{AB}}\otimes|00\rangle_{A'B'})\\
          &=\Phi(\bigoplus_{i, j} \sqrt{q_{i j}}[c_{1z}\cdot(|2i\rangle\langle2i|-|2i+1\rangle\langle2i+1|)+c_{1x}\cdot(|2i\rangle\langle2i+1|+|2i+1\rangle\langle2i|)]\\&\qquad(\cos (\frac{\theta}{2})|2i,2j\rangle-\sin (\frac{\theta}{2})|2i+1,2j+1\rangle)_{AB}\otimes|00\rangle_{A'B'})\\
          &=\bigoplus_{i, j}\sqrt{q_{i j}}|2i,2j\rangle_{AB}\otimes (c_{1z}\cdot  Z+c_{1x}\cdot  X)(\cos (\frac{\theta}{2})|00\rangle_{A'B'}-\sin(\frac{\theta}{2})|11\rangle_{A'B'})\\
          &=|\sigma\rangle_{AB}\otimes \textcolor{black}{{A_0}|\psi_{\theta}\rangle_{A' B'}}
        \end{split}
        \end{equation}
         where $|\sigma\rangle_{AB}$ is a bipartite `junk' state.
        Analogous self-testing equations hold for ${B_1}$, ${B_0}$ as well as  ${A_x}{B_y}$ for $x, y \in \{0,1\}$.
        
        
        
        \end{itemize}
\end{proof}
\section{The guessing probability of a quantum adversary under Colored Noise}\label{randomness under quantum adversary}

As a quantum adversary, Eve holds a quantum system that may be entangled with the device held by Alice and Bob. Let us denote the tripartite quantum state of Alice, Bob and Eve by $|\Psi \rangle_{ABE}$. 
In this section, we will calculate the guessing probability of such an adversary Eve, when Alice and Bob perform the tilted Hardy test on their reduced system $\rho_{AB}$ and observe that the three Hardy zero constraints are perfectly satisfied while the Hardy probability achieves some non-maximal value $0 < p \leq P_{Hardy}^{w=w_0,Q} = \frac{(4 w_0+5)\sqrt{4 w_0+5}-(12 w_0 +11)}{2( w_0 +1)}$. Here we fix $w = w_0$, in which case as discussed in Section~\ref{tilted_hardy}, one can certify the most global randomness from the measurement setting $(A_0, B_0)$, when the optimal quantum behaviour is observed. The problem we want to solve is to maximize the guessing probability under the constraint that (i) Alice, Bob and Eve's system is quantum, (ii) the reduced system of Alice and Bob satisfies the three Hardy zero constraints, and (iii) the reduced system of Alice and Bob achieves some value $0 < p \leq P_{Hardy}^{w=w_0,Q}$ for the Hardy probability, i.e., 
\begin{equation}
  \begin{split}
    \text{max} \; &\textcolor{black}{P_{guess}(A,B|A_0,B_0,E)}\\
     s.t.\;~
           &P_{A,B|X,Y}(0,1|A_0,B_1)=P_{A,B|X,Y}(1,0|A_1,B_0)=P_{A,B|X,Y}(0,0|A_1,B_1)=0,\\
           &P_{A,B|X,Y}(0,0|A_0,B_0) + w_0 P_{A,B|X,Y}(1,1|A_0,B_0) = p,\\
           & P_{A,B,E|X,Y,Z} \in \mathcal{Q}, 
  \end{split}
\end{equation}
where $\mathcal{Q}$ denotes the set of three-party quantum boxes. 

When the three zero constraints are satisfied, solving Eq.~\eqref{hardy_value_vio} for $P_{A,B|X,Y}(0,0|A_0,B_0) + w_0 P_{A,B|X,Y}(1,1|A_0,B_0) = p$ gives rise to two different values of $\theta$, namely
\begin{equation}\label{two_solution}
\begin{split}
\theta_1^{p,w_0}=&\arcsin\left(\frac{1-p}{3}+\frac{2}{3}\sqrt{1+10p+p^2-12w_0}\sin\left[\frac{\pi}{6}\right.\right.\\&\left.\left.-\frac{1}{3}\arccos\left(\frac{p^3+15p^2+p(39-18w_0)-36w_0-1}{\sqrt{(1+10p+p^2-12w_0)^3}}\right)\right]\right),\\
\theta_2^{p,w_0}=&\arcsin\left(\frac{1-p}{3}+\frac{2}{3}\sqrt{1+10p+p^2-12w_0}\sin\left[\frac{\pi}{6}\right.\right.\\&\left.\left.+\frac{1}{3}\arccos\left(\frac{p^3+15p^2+p(39-18w_0)-36w_0-1}{\sqrt{(1+10p+p^2-12w_0)^3}}\right)\right]\right).
\end{split}
\end{equation}
Importantly, the measurements on Alice and Bob's system for the two corresponding states $| \psi_{\theta} \rangle$ are also fixed (and different) as specified in Eq.~\eqref{eq:w-measurements} so that the zero constraints cannot be satisfied by mixed states on Alice and Bob's system. 

\begin{figure}[H]
    \centering
    \includegraphics[width=0.7\textwidth]{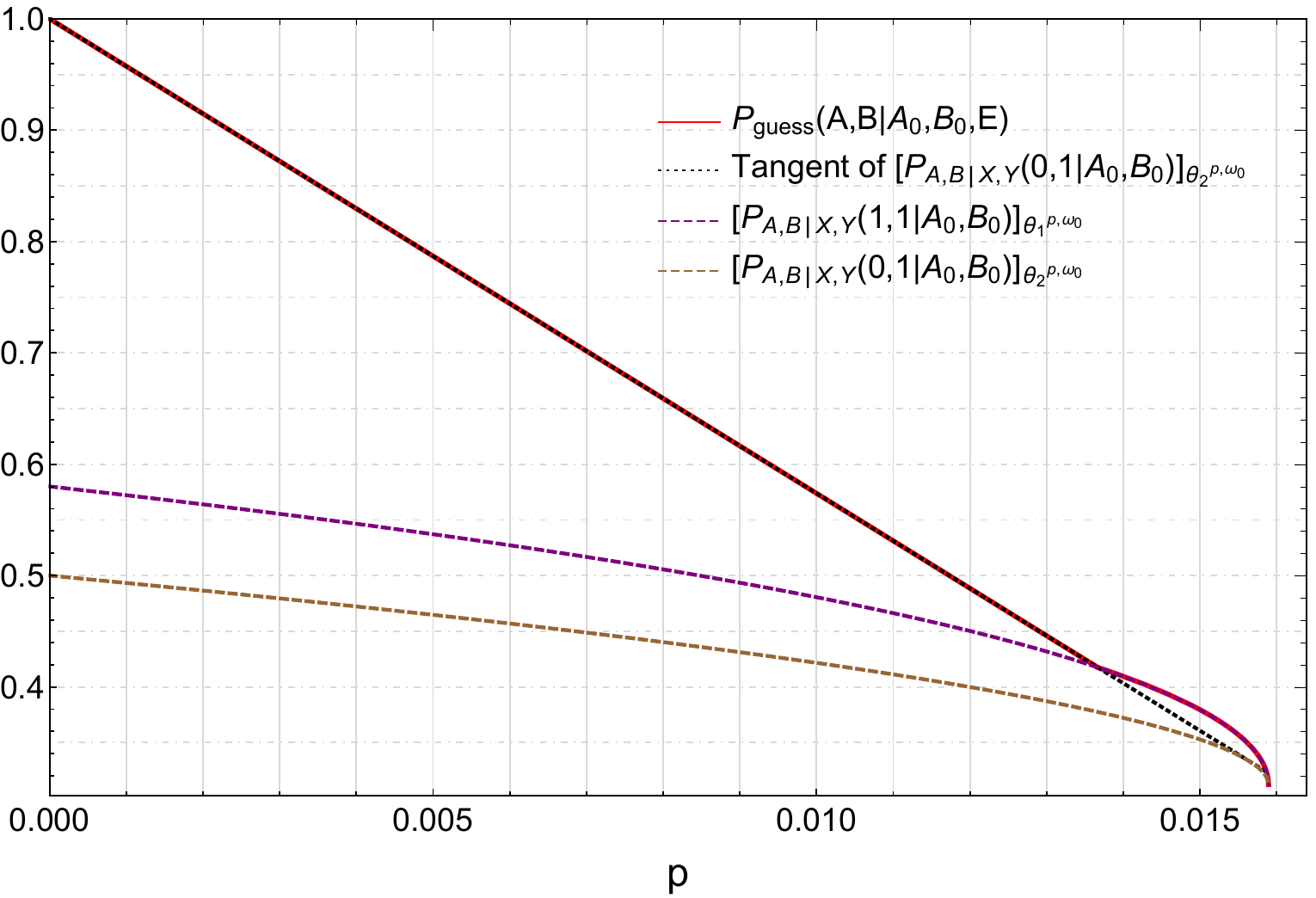}
    \caption{$P_{guess}(A,B|A_0,B_0,E)$,$\left[P_{A,B|X,Y}(1,1|A_0,B_0) \right]_{\theta_1^{p,w_0}}$, $\left[P_{A,B|X,Y}(0,1|A_0,B_0) \right]_{\theta_2^{p,w_0}}$ and the tangent of  $\left[P_{A,B|X,Y}(0,1|A_0,B_0) \right]_{\theta_2^{p,w_0}}$ versus the Hardy probability $p$. The solid red line denotes the guessing probability $P_{guess}(A,B|A_0,B_0,E)$. The dotted black line denotes the tangent of the probability $\left[P_{A,B|X,Y}(0,1|A_0,B_0) \right]_{\theta_2^{p,w_0}}$. The dashed purple and brown lines denote the probabilities $\left[P_{A,B|X,Y}(1,1|A_0,B_0) \right]_{\theta_1^{p,w_0}}$, $\left[P_{A,B|X,Y}(0,1|A_0,B_0) \right]_{\theta_2^{p,w_0}}$, respectively.}
    \label{Guessing Prob for Hardy value}
\end{figure}

Therefore, Eve's strategy in this case just consists of a mixture of extreme quantum boxes (for which the state of Alice and Bob's system is pure and decoupled from that of Eve) under the constraint that the observed Hardy probability is $p$. For any extreme quantum behavior $\{ P_{A,B|X,Y} \}_{\theta^{p,w_0}}$ parameterized by the value $\theta^{p,w_0}$, one can readily calculate the guessing probability directly from the behavior as
\begin{equation}
\begin{split}
\label{eq:pguess}
    \left[\textcolor{black}{P_{guess}(A,B|A_0,B_0,E)} \right]_{\theta^{p,w_0}} &=\max_{\{a,b\}}  \left[ P_{A,B|X,Y}(a,b|A_0,B_0) \right]_{\theta^{p,w_0}}\\&=\left\{
\begin{array}{ll}
\left[P_{A,B|X,Y}(1,1|A_0,B_0) \right]_{\theta^{p,w_0}},& \; \;  {0<\theta^{p,w_0} \leq \theta^{w_0}}\\
\left[P_{A,B|X,Y}(0,1|A_0,B_0) \right]_{\theta^{p,w_0}}, & \; \;  { \theta^{w_0}<\theta^{p,w_0} \leq\frac{\pi}{2}}\\
\end{array} \right.\\
\end{split}
\end{equation}
where $\theta^{w_0}= \arcsin \left(-4\left(\frac{2}{3(9+\sqrt{177})}\right)^{\frac{1}{3}}+\frac{2}{3}^{\frac{2}{3}}\left(9+\sqrt{177}\right)^{\frac{1}{3}} \right)\approx 1.1356$ which corresponds to the optimal quantum behaviour $P_{Hardy}^{w_0,Q}$. Here 
$\left[P_{A,B|X,Y}(0,1|A_0,B_0) \right]_{\theta} =\linebreak \left[P_{A,B|X,Y}(1,0|A_0,B_0) \right]_{\theta} = \frac{\sin^3\theta}{2(2-\sin\theta)^2}$ and $\left[P_{A,B|X,Y}(1,1|A_0,B_0) \right]_{\theta} = \frac{1}{2}+ \frac{\cos\theta(2+\sin\theta)(1-\sin\theta)^{\frac{1}{2}}}{2(2-\sin\theta)(1+\sin\theta)^{\frac{1}{2}}}-\frac{\sin^3\theta}{2(2-\sin\theta)^2}$. Note that $\theta_1^{p,w_0} < \theta^{w_0} < \theta_2^{p,w_0}$ for all $p \in (0, P_{Hardy}^{w=w_0,Q})$ and the corresponding guessing probabilities for these two extreme boxes are related as $\left[P_{A,B|X,Y}(1,1|A_0,B_0) \right]_{\theta_1^{p,w_0}}\geq \left[ P_{A,B|X,Y}(0,1|A_0,B_0) \right]_{\theta_2^{p,w_0}}$, $\forall p\in (0,P_{Hardy}^{w_0,Q}]$. Furthermore, as a function of $p$, the probabilities $\left[P_{A,B|X,Y}(1,1|A_0,B_0) \right]_{\theta_1^{p,w_0}}$ and $\left[P_{A,B|X,Y}(0,1|A_0,B_0) \right]_{\theta_2^{p,w_0}}$ are both concave and monotonically decreasing (see Fig.~\ref{Guessing Prob for Hardy value}).  

The general strategy of any adversary Eve is thus to provide Alice and Bob with (a mixture of) extreme quantum boxes $\{ P_{A,B|X,Y} \}$ that achieve the Hardy probability $p$. To do this, she may admix some local deterministic boxes (which allow a guessing probability of $1$) with quantum behaviors that achieve a Hardy probability in the range $(p, P_{Hardy}^{w_0, Q}]$. Solving for the optimal admixture of extreme boxes for which the guessing probability is given by the equation \eqref{eq:pguess} with local boxes from the Tables~\ref{table_hardy1} and \ref{table_hardy2} gives the guessing probability for any given $p\in[0,P_{Hardy}^{w_0,Q}]$.

From Appendix~\ref{proof-w-hardy}, we know there are totally $5$ local hidden variable deterministic behaviours which satisfy the three Hardy zero constraints. The local boxes shown in Table~\ref{table_hardy1} satisfy $\left[P_{A,B|X,Y}(1,1|A_0,B_0)\right]_{\text{LHV}_I}=1$, and these boxes achieve a Hardy probability $\left[P_{A,B|X,Y}(0,0|A_0,B_0) + w_0 P_{A,B|X,Y}(1,1|A_0,B_0) \right]_{\text{LHV}_I} =w_0<0$. Thus, when Eve chooses a probabilistic mixture of a deterministic box from Table~\ref{table_hardy1} and any pure quantum behaviour ($\theta_1^{\textcolor{black}{\widetilde{p}},w_0}$) with $p<\textcolor{black}{\widetilde{p}}\leq P_{Hardy}^{w_0,Q}$, the following equation holds (from the constraint that $P_{Hardy} = p$):
\begin{equation}
    r\cdot \textcolor{black}{\widetilde{p}}+(1-r)\cdot w_0=p \quad \Rightarrow \quad r=\frac{p-w_0}{\textcolor{black}{\widetilde{p}}-w_0}.
\end{equation}
Such a mixture can not increase the guessing probability, due to the fact that
\begin{equation}
\begin{split}
    \frac{p-w_0}{\textcolor{black}{\widetilde{p}}-w_0}\cdot \left[P_{A,B|X,Y}(1,1|A_0,B_0)\right]_{\theta_1^{\textcolor{black}{\widetilde{p}},w_0}} &+\left(1-\frac{p-w_0}{\textcolor{black}{\widetilde{p}}-w_0}\right)\cdot1\leq \left[P_{A,B|X,Y}(1,1|A_0,B_0)\right]_{\theta_1^{p,w_0}}\quad 
    \\ & \forall p,\textcolor{black}{\widetilde{p}}\text{ with } 0\leq p<\textcolor{black}{\widetilde{p}}\leq P_{Hardy}^{w_0,Q}.
    \end{split}
    \end{equation}
Therefore, the admixture of local deterministic boxes from Table~\ref{table_hardy1} does not increase the guessing probability.

    

On the other hand, any local deterministic box from Table~\ref{table_hardy2} satisfies \linebreak $\left[P_{A,B|X,Y}(0,1|A_0,B_0)\right]_{\text{LHV}_{II}}=1$ and achieves the zero Hardy probability \linebreak$\left[P_{A,B|X,Y}(0,0|A_0,B_0) + w_0 P_{A,B|X,Y}(1,1|A_0,B_0) \right]_{\text{LHV}_{II}}=0$. When Eve chooses a probabilistic mixture of a deterministic box from Table~\ref{table_hardy2} and any pure quantum behavior ($\theta_2^{\textcolor{black}{\widetilde{p}},w_0}$) with $p<\textcolor{black}{\widetilde{p}}\leq P_{Hardy}^{w_0,Q}$, the following equation holds (from the constraint that $P_{Hardy} = p$)  
\begin{equation}
    r\cdot \textcolor{black}{\widetilde{p}}+(1-r)\cdot0=p \quad \Rightarrow \quad r=\frac{p}{\textcolor{black}{\widetilde{p}}}
\end{equation}
We can find the optimal $\textcolor{black}{\widetilde{p}}$ which maximizes the guessing probability of Eve:
\begin{equation}
  \begin{split}
   \max_{\textcolor{black}{\widetilde{p}}}&\quad \frac{p}{\textcolor{black}{\widetilde{p}}} \cdot \left[P_{A,B|X,Y}(0,1|A_0,B_0)\right]_{\theta_2^{\textcolor{black}{\widetilde{p}},w_0}}+\left(1-\frac{p}{\textcolor{black}{\widetilde{p}}}\right)\cdot 1\\
    s.t. & \quad p<\textcolor{black}{\widetilde{p}}\leq P_{Hardy}^{w_0,Q},\\
    &\quad \forall p\in (0,P_{Hardy}^{w_0,Q}).\\
  \end{split}
\end{equation}
The above problem is equivalent to find the $\textcolor{black}{\widetilde{p}}$ on the curve of the function 
\linebreak\textcolor{black}{$\left[P_{A,B|X,Y}(0,1|A_0,B_0)\right]_{\theta_{2}^{p,w_0}}$}
(as the function of $p$) where the tangent intersects the deterministic strategy point  $(0,1)$. That means
\begin{equation}
   \frac{\partial{\left[P_{A,B|X,Y}(0,1|A_0,B_0)\right]_{\theta_{2}^{p,w_0}}}}{\partial p}\Biggr|_{\widetilde{p}} \cdot\widetilde{p}+1=\left[P_{A,B|X,Y}(0,1|A_0,B_0)\right]_{\theta_{2}^{\widetilde{p},w_0}}
\end{equation}
The solution of the above optimization on $\widetilde{p}$ is denoted as $\widetilde{p^{*}}\approx0.01563$. From the results, we obtain the tangent curve of $\left[P_{A,B|X,Y}(0,1|A_0,B_0)\right]_{\theta_{2}^{p,w_0}}$
(as the function of $p$),
\begin{equation}
    \frac{(\left[P_{A,B|X,Y}(0,1|A_0,B_0)\right]_{\theta_{2}^{\widetilde{p^{*}}},w_0})-1}{\widetilde{p^{*}}}\cdot p+1
\end{equation}

As shown in Fig.~\ref{Guessing Prob for Hardy value}, the relation  $\left[P_{A,B|X,Y}(1,1|A_0,B_0) \right]_{\theta_1^{p,w_0}}\geq \left[ P_{A,B|X,Y}(0,1|A_0,B_0) \right]_{\theta_2^{p,w_0}}$ holds for $ p\in (0,P_{Hardy}^{w_0,Q}]$. The tangent curve of $\left[P_{A,B|X,Y}(0,1|A_0,B_0)\right]_{\theta_{2}^{p,w_0}}$ must intersect the curve of $\left[P_{A,B|X,Y}(1,1|A_0,B_0) \right]_{\theta_1^{p,w_0}}$ at a certain point for which we denote the horizontal ordinate as $p^{*}$. Therefore, in any Eve's mixed strategy for the value $p\in[0,P_{Hardy}^{w_0,Q}]$, Eve's guessing probability is given as

\begin{equation}
\begin{split}
&P_{guess}(A,B|A_0,B_0,E)=\left\{
\begin{array}{ll}
\frac{(\left[P_{A,B|X,Y}(0,1|A_0,B_0)\right]_{\theta_{2}^{\widetilde{p^{*}},w_0}})-1}{\widetilde{p^{*}}}\cdot p+1 \quad & 0\leq p\leq p^{*}\\
\left[ P_{A,B|X,Y}(1,1|A_0,B_0) \right]_{\theta_1^{p,w_0}} \quad & p^{*}< p\leq P_{Hardy}^{w_0,Q}\\
\end{array} \right.\\
\end{split}
\end{equation}
where $p^{*} \approx 0.01366$. 

\section{$\alpha$-CHSH and its robust self-testing statement  }\label{robust_self_testing_App}
In this section, We first introduce a family of bipartite Bell expressions named $\alpha$-CHSH (used in Eq.~\eqref{Bell_ineq1}) with a parameter $\alpha>\frac{1}{3}$:
\begin{equation}
    I_{\alpha}=\alpha A_0B_0+A_0B_1+A_1B_0-A_1B_1
\end{equation}
where $A_0,A_1$ are two binary observables for Alice and $B_0,B_1$ are two binary observables for Bob.

The maximum value of $I_{\alpha}$ in quantum theory is $I_{\alpha}^Q=(\alpha+1)\sqrt{\frac{\alpha+1}{\alpha}}$, and it's achieved by the specific reference quantum system (up to a local isometry) :
\begin{equation}
    \begin{split}
        &\textcolor{black}{|\phi^+\rangle}=\frac{1}{\sqrt{2}}(|00\rangle +|11\rangle)\\
        &A_0=\sigma_z,\quad\qquad\qquad\qquad\qquad\qquad\qquad \qquad 
        A_1=\frac{\alpha-1}{2\alpha}\sigma_z+\frac{\sqrt{(\alpha+1)(3\alpha-1)}}{2\alpha}\sigma_x\\
        &B_0=\frac{2\alpha-1}{2\alpha}\sqrt{\frac{\alpha+1}{\alpha}}\sigma_z+\frac{1}{2\alpha}\sqrt{\frac{3\alpha-1}{\alpha}}\sigma_x, \quad 
        B_1=\frac{1}{2}\sqrt{\frac{\alpha+1}{\alpha}}\sigma_z-\frac{1}{2}\sqrt{\frac{3\alpha-1}{\alpha}}\sigma_x\\
    \end{split}
\end{equation}
The maximum value of $I_{\alpha}$ in classical theory is easily verified to be:
 \begin{equation}
        I_{\alpha}^{LHV}\leq \left\{
        \begin{array}{cc}
        3-\alpha,     & \qquad \frac{1}{3}<\alpha\leq1 \\
        1+\alpha,     & \qquad 1<\alpha
        \end{array}
        \right.
    \end{equation}

Let define the shifted Bell operator as $\overline{I_{\alpha}}:=I_{\alpha}^Q\mathbb{I}-I_{\alpha}=I_{\alpha}^Q\mathbb{I}-\alpha A_0B_0-A_0B_1-A_1B_0+A_1B_1$, 
 and take a trivial monomials basis $R=[R_1=A_0,R_2=A_1,R_3=B_0,R_4=B_1]^T$ to do the sum of squares (SOS) decomposition of the shifted operator Bell $\overline{I_{\alpha}}=R^{\dag}M_{\alpha}R$. Obviously, $\overline{I_{\alpha}}$ is inherently positive semidefinite,  and we get:
\begin{equation}
    M_{\alpha}=\left[\begin{array}{cccc}
\frac{\sqrt{\alpha(\alpha+1)}}{2} & 0 & -\frac{\alpha}{2} & -\frac{1}{2} \\
0 & \frac{1}{2}\sqrt{\frac{\alpha+1}{\alpha}}& -\frac{1}{2} & \frac{1}{2} \\
-\frac{\alpha}{2} & -\frac{1}{2} & \frac{\sqrt{\alpha(\alpha+1)}}{2} & 0\\
-\frac{1}{2} & \frac{1}{2} & 0 & \frac{1}{2}\sqrt{\frac{\alpha+1}{\alpha}}
\end{array}\right]\geq0
\end{equation}
One can do Cholesky Decomposition of $M_{\alpha}$ to get two polynomials $P_1,P_2$ and the shifted Bell operator is expressed as $\overline{I_{\alpha}}=P_1^{\dag}P_1+P_2^{\dag}P_2$.
\begin{equation}\label{sos}
    \begin{split}
        P_1&=\frac{[\alpha(1+\alpha)]^{\frac{1}{4}}}{\sqrt{2}}A_0-\frac{\alpha^{^{\frac{3}{4}}}}{\sqrt{2}(1+\alpha)^{\frac{1}{4}}}B_0-\frac{1}{\sqrt{2}[\alpha(1+\alpha)]^{\frac{1}{4}}}B_1,\\
        P_2&=\frac{(1+\alpha)^{\frac{1}{4}}}{\sqrt{2}\alpha^{\frac{1}{4}}}A_1-\frac{\alpha^{^{\frac{1}{4}}}}{\sqrt{2}(1+\alpha)^{\frac{1}{4}}}B_0+\frac{\alpha^{\frac{1}{4}}}{\sqrt{2}(1+\alpha)^{\frac{1}{4}}}B_1.\\
    \end{split}
\end{equation}

Due to the experimental noise, usually the ideal maximum quantum value $I_{\alpha}^Q$ cannot be observed in the experiments but up to some error term $e$. To distinguish the notations used by the reference quantum system above, we will refer to the physical state and observables (hermitian and dichotomic) to prime notations $|\psi'\rangle,A'_0,A'_1,B'_0,B'_1$.
\begin{figure}[H]
    \centering
    \includegraphics[width=0.7\textwidth]{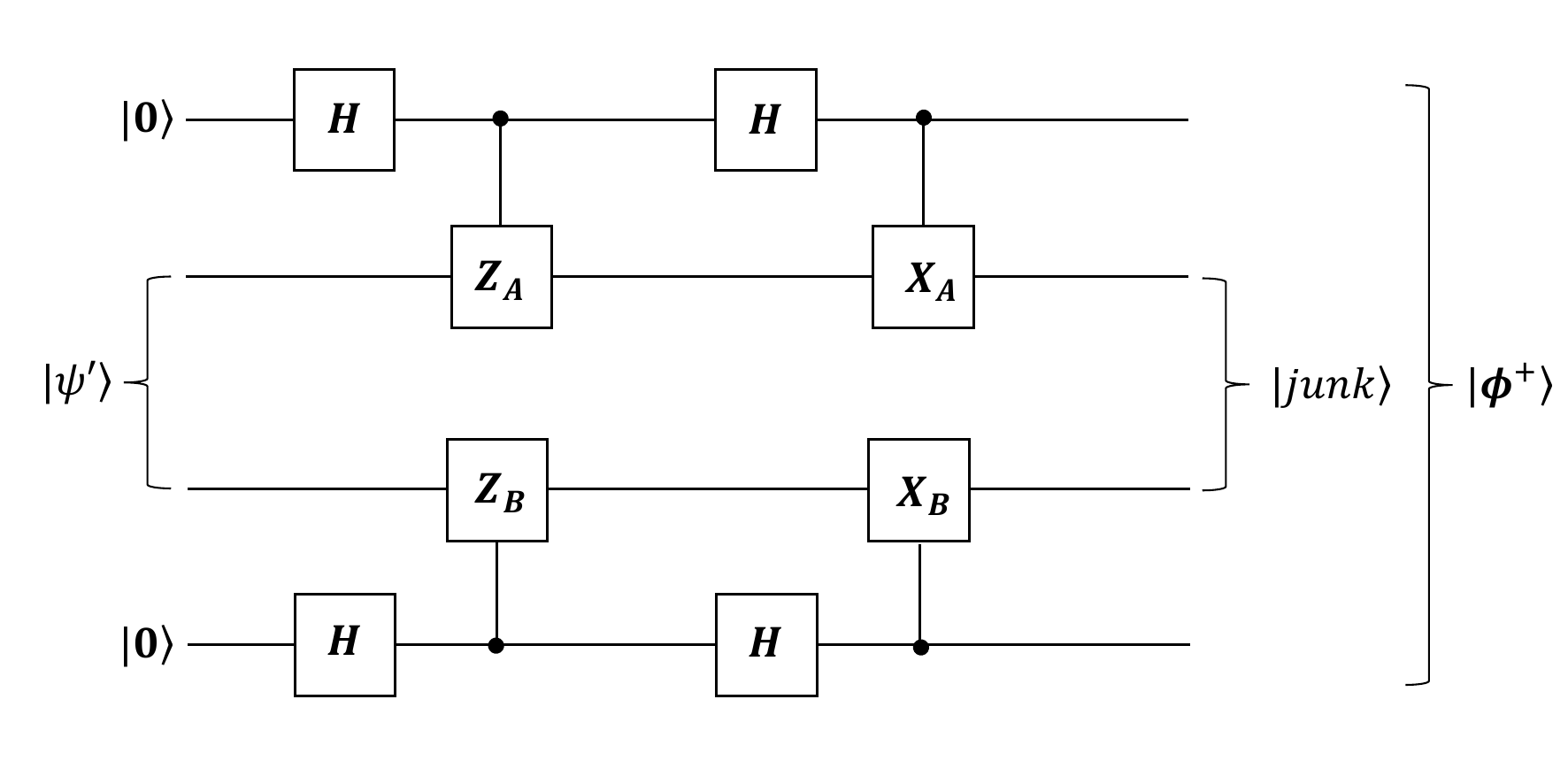}
    \caption{The local isometry $\Phi=\Phi_A\otimes \Phi_B$ used to self-test the reference state and measurements. }
    \label{swap}
\end{figure}

This Bell expression $I_{\alpha}$ provides a robust self-test for its reference state $\textcolor{black}{|\phi^+\rangle}$ and reference observables $A_0,A_1,B_0,B_1$, to prove this robust self-testing statement, we show there exist a local isometry $\Phi=\Phi_A\otimes \Phi_B$, as shown in Fig.~\ref{swap} acting on the physical system and its output state is close to the corresponding reference system in a noise-tolerant way (the error term is of order $O(\sqrt{e})$) with some $|junk\rangle$ state. More precisely:
\begin{equation}
     \left\|\Phi\left(A'_x\otimes B'_y|\psi'\rangle\right)-\left(A_x\otimes B_y\textcolor{black}{|\phi^+\rangle}\right)\otimes|junk\rangle\right\| \leq O(\sqrt{e})
\end{equation}
where $x,y\in\{-1,0,1\}$, the subscript -1 refers to $\mathbb{I}$ and the operators $X_A,X_B,Z_A,Z_B$ in the isometry are defined in terms of $A'_x,B'_y$, $x,y\in\{0,1\}$.

\begin{proof}
The proof of the robust self-testing statement includes the following 4 steps.

\begin{enumerate}
    \item [Step. 1] \textbf{Defining the operators in the isometry.}
    
Firstly, we define the following four local operators (with the prime notation) in terms of $A'_x,B'_y$, $x,y\in\{0,1\}$:
\begin{equation}
    \begin{split}
        &X'_A=\frac{2\alpha}{(\alpha+1)^{\frac{1}{2}}(3\alpha-1)^{\frac{1}{2}}}A'_1-\frac{\alpha-1}{(\alpha+1)^{\frac{1}{2}}(3\alpha-1)^{\frac{1}{2}}}A'_0,\quad \quad
        Z'_A=A'_0.\\
        &X'_B=\frac{\alpha^{\frac{1}{2}}}{(3\alpha-1)^{\frac{1}{2}}}B'_0-\frac{2\alpha-1}{\alpha^{\frac{1}{2}} (3\alpha-1)^{\frac{1}{2}}}B'_1,\quad
        Z'_B=\frac{\alpha^{\frac{1}{2}}}{(\alpha+1)^{\frac{1}{2}}}B'_0+\frac{1}{\alpha^{\frac{1}{2}}(\alpha+1)^{\frac{1}{2}}}B'_1.\\
    \end{split}
\end{equation}

In order to make the operators in the isometry unitary, we play the regularisation trick \cite{bamps2015sum} on these operators. For operator $X'_A$, we first replace the zero eigenvalues into 1  and denote the new operator as $X^*_A$, then normalize the eigenvalues by $X_A=\frac{X^*_A}{|X^*_A|}$, such that the resulting operator $X_A$ is Hermitian and unitary and has the property $X_A=\frac{X'_A}{|X'_A|}$. Then play the same regularisation trick to define $X_B, Z_A, Z_B$. Note that $Z'_A=A'_0$ is already unitary, so $Z_A=Z'_A$.

\item [Step. 2] \textbf{Deriving some useful bounds for physical operators.}
 
According to the SOS decomposition \eqref{sos}, suppose the observation of the quantum value is  $\langle I_{\alpha}\rangle\geq I_{\alpha}^Q-e$, i.e., $\langle \overline{I_{\alpha}}\rangle=\|P'_1|\psi'\rangle\|^2+\|P'_2|\psi'\rangle\|^2\leq e$, we have the following two bounds:
\begin{equation}\label{p1R}
    \left\|P'_1|\psi'\rangle\right\|=\left\|\frac{1}{\sqrt{2}\alpha^{\frac{1}{4}}(1+\alpha)^{\frac{1}{4}}}\left[\alpha^{\frac{1}{2}}(1+\alpha)^{\frac{1}{2}}A'_0-\alpha B'_0-B'_1\right]|\psi'\rangle\right\|\leq\sqrt{e}
\end{equation}
\begin{equation}\label{p2R}
    \left\|P'_2|\psi'\rangle\right\|=\left\|\frac{1}{\sqrt{2}\alpha^{\frac{3}{4}}(1+\alpha)^{\frac{1}{4}}}\left[\alpha^{\frac{1}{2}}(1+\alpha)^{\frac{1}{2}}A'_1-\alpha B'_0+\alpha B'_1\right]|\psi'\rangle\right\|\leq\sqrt{e}
\end{equation}

From \eqref{p1R} and \eqref{p2R}, directly we get:
\begin{equation}\label{zEquR}
    \left\|\left(Z'_A-Z'_B\right)|\psi'\rangle\right\|=\left\|\frac{\sqrt{2}\alpha^{\frac{1}{4}}(1+\alpha)^{\frac{1}{4}}}{\alpha^{\frac{1}{2}}(1+\alpha)^{\frac{1}{2}}}P'_1|\psi'\rangle\right\|\leq \frac{\sqrt{2}}{\alpha^{\frac{1}{4}}(1+\alpha)^{\frac{1}{4}}}\sqrt{e}\equiv\delta_1
\end{equation}

\begin{equation}\label{xEquR}
\begin{split}
    \left\|\left(X'_A-X'_B\right)|\psi'\rangle\right\|&=\left\|\left[\frac{2\sqrt{2}\alpha^{\frac{5}{4}}}{(3\alpha-1)^{\frac{1}{2}}(1+\alpha)^{\frac{3}{4}}}P'_2-\frac{\sqrt{2}(\alpha-1)}{\alpha^{\frac{1}{4}}(3\alpha-1)^{\frac{1}{2}}(1+\alpha)^{\frac{3}{4}}}P'_1\right]|\psi'\rangle\right\|\\
    &\leq \frac{2\sqrt{2}\alpha^{\frac{3}{2}}+\sqrt{2}|\alpha-1|}{\alpha^{\frac{1}{4}}(3\alpha-1)^{\frac{1}{2}}(1+\alpha)^{\frac{3}{4}}}\sqrt{e}\equiv\delta_2
    \end{split}
\end{equation}
To derive the bounds for anti-coummuting operators $\left\|\{X'_A,Z'_A\}|\psi'\rangle\right\|$ and $\left\|\{X'_B,Z'_B\}|\psi'\rangle\right\|$, we will need the following relations:
\begin{equation}
    \begin{split}
      {X'_A}^2&=\left[\frac{2\alpha A'_1-(\alpha-1)A'_0}{(\alpha+1)^{\frac{1}{2}}(3\alpha-1)^{\frac{1}{2}}}\right]^2=\frac{4\alpha^2+(\alpha-1)^2}{(\alpha+1)(3\alpha-1)}\mathbb{I}-\frac{2\alpha(\alpha-1)}{(\alpha+1)(3\alpha-1)}\{A'_0,A'_1\},\\
      {Z'_A}^2&={A'_0}^2=\mathbb{I},\\
      {X'_B}^2&=\left[\frac{\alpha B'_0-(2\alpha-1)B'_1}{\alpha^{\frac{1}{2}}(3\alpha-1)^{\frac{1}{2}}}\right]^2=\frac{\alpha^2+(2\alpha-1)^2}{\alpha(3\alpha-1)}\mathbb{I}-\frac{2\alpha-1}{3\alpha-1}\{B'_0,B'_1\},\\
      {Z'_B}^2&=\left[\frac{\alpha B'_0+B'_1}{\alpha^{\frac{1}{2}}(\alpha+1)^{\frac{1}{2}}}\right]^2=\frac{\alpha^2+1}{\alpha(\alpha+1)}\mathbb{I}+\frac{1}{\alpha+1}\{B'_0,B'_1\}.\\
    \end{split}
\end{equation}
Thus, the following equations holds true:
{\small
\begin{equation}\label{antiComUP}
\begin{split}
        &\left\|\left[\{X'_A,Z'_A\}-\{X'_B,Z'_B\}\right]|\psi'\rangle\right\|\\&\leq \left\| X'_A(Z'_A-Z'_B)|\psi'\rangle\right\|+\left\| X'_B(Z'_A-Z'_B)|\psi'\rangle\right\|+\left\| Z'_A(X'_A-X'_B)|\psi'\rangle\right\|+\left\| Z'_B(X'_A-X'_B)|\psi'\rangle\right\|\\
        &=\left\|(X'_A-X'_B)|\psi'\rangle\right\|+\sqrt{\langle\psi'|(Z'_A-Z'_B){X'_A}^2(Z'_A-Z'_B)|\psi'\rangle}\\
        &\quad+\sqrt{\langle\psi'|(Z'_A-Z'_B){X'_B}^2(Z'_A-Z'_B)|\psi'\rangle}+\sqrt{\langle\psi'|(X'_A-X'_B){Z'_B}^2(X'_A-X'_B)|\psi'\rangle}\\
        &\leq\delta_2+\sqrt{\frac{4\alpha^2+(\alpha-1)^2}{(\alpha+1)(3\alpha-1)}\delta_1^2+\frac{2\alpha|\alpha-1|}{(\alpha+1)(3\alpha-1)}\left\|(A'_0-A'_1)(Z'_A-Z'_B)|\psi'\rangle\right\|^2}\\
        &\quad+\sqrt{\frac{\alpha^2+(2\alpha-1)^2}{\alpha(3\alpha-1)}\delta_1^2+\frac{|2\alpha-1|}{3\alpha-1}\left\|(B'_0-B'_1)(Z'_A-Z'_B)|\psi'\rangle\right\|^2}\\&\quad+\sqrt{\frac{\alpha^2+1}{\alpha(\alpha+1)}\delta_2^2+\frac{1}{\alpha+1}\left\|(B'_0+B'_1)(X'_A-X'_B)|\psi'\rangle\right\|^2}\\
        &\leq\delta_2+\sqrt{\frac{4\alpha^2+(\alpha-1)^2}{(\alpha+1)(3\alpha-1)}+\frac{8\alpha|\alpha-1|}{(\alpha+1)(3\alpha-1)}}\delta_1\\
        &\quad+\sqrt{\frac{\alpha^2+(2\alpha-1)^2}{\alpha(3\alpha-1)}+\frac{4|2\alpha-1|}{3\alpha-1}}\delta_1+\sqrt{\frac{\alpha^2+1}{\alpha(\alpha+1)}+\frac{4}{\alpha+1}}\delta_2\\&\equiv\delta_3\\
\end{split}
\end{equation}}

The left side of \eqref{antiComUP} can also be expressed as:
{\small
\begin{equation}\label{antiComLOW}
\begin{split}
&\left\|\left[\{X'_A,Z'_A\}-\{X'_B,Z'_B\}\right]|\psi'\rangle\right\|\\&=\left\| \left[\frac{2}{(1+\alpha)^{\frac{1}{2}}(3\alpha-1)^{\frac{1}{2}}}\left[\alpha\{A'_0,A'_1\}-(1-\alpha)\{B'_0,B'_1\}\right]-\frac{2(\alpha-1)(2\alpha-1)}{\alpha(3\alpha-1)^{\frac{1}{2}}(1+\alpha)^{\frac{1}{2}}}\mathbb{I}\right]|\psi'\rangle\right\|\\
&=\left\| \left[
\frac{2\sqrt{2}\alpha^{\frac{3}{4}}}{(1+\alpha)^{\frac{3}{4}}(3\alpha-1)^{\frac{1}{2}}}(A'_1P'_1+\alpha^{\frac{1}{2}}A'_0P'_2)+\frac{2\sqrt{2}\alpha^{\frac{3}{4}}}{(1+\alpha)^{\frac{5}{4}}(3\alpha-1)^{\frac{1}{2}}}(\alpha B'_0P'_2+B'_1P'_2
+\alpha^{\frac{1}{2}}B'_0P'_1-\alpha^{\frac{1}{2}}B'_1P'_1)\right.\right.\\
&\left.\left. +\frac{2(\alpha-1)}{(1+\alpha)^{\frac{3}{2}}(3\alpha-1)^{\frac{1}{2}}}\{B'_0,B'_1\}-\frac{2(\alpha-1)^2}{\alpha(1+\alpha)^{\frac{3}{2}}(3\alpha-1)^{\frac{1}{2}}}\mathbb{I} \right]|\psi'\rangle\right\|\\
&\geq \frac{2|\alpha-1|}{(1+\alpha)^{\frac{3}{2}}(3\alpha-1)^{\frac{1}{2}}}\left\|\left(\{B'_0,B'_1\}-\frac{\alpha-1}{\alpha}\mathbb{I}\right)|\psi'\rangle\right\|-\frac{2\sqrt{2}\alpha^{\frac{3}{4}}(1+\alpha^{\frac{1}{2}})\left[(1+\alpha)^{\frac{1}{2}}+(1+\alpha^{\frac{1}{2}})\right]}{(1+\alpha)^{\frac{5}{4}}(3\alpha-1)^{\frac{1}{2}}}\sqrt{e}
\end{split}
\end{equation}}

From the above two inequalities \eqref{antiComUP} and \eqref{antiComLOW}, we have:
{\small
\begin{equation}\label{Bbound}
\begin{split}
\left\|\left(\{B'_0,B'_1\}-\frac{\alpha-1}{\alpha}\mathbb{I}\right)|\psi'\rangle\right\|&\leq\frac{(1+\alpha)^{\frac{3}{2}}(3\alpha-1)^{\frac{1}{2}}}{2|\alpha-1|}\delta_3\\
&\quad+\frac{\sqrt{2}\alpha^{\frac{3}{4}}(1+\alpha^{\frac{1}{2}})(1+\alpha)^{\frac{1}{4}}\left[(1+\alpha)^{\frac{1}{2}}+(1+\alpha^{\frac{1}{2}})\right]}{|\alpha-1|}\sqrt{e}\\&\equiv\delta_4
\end{split}
\end{equation}}
Then we can get:
\begin{equation}\label{anticomR}
\begin{split}
    \left\|\{X'_B,Z'_B\}|\psi'\rangle\right\|&=\frac{2|\alpha-1|}{(3\alpha-1)^{\frac{1}{2}}(\alpha+1)^{\frac{1}{2}}}\left\|\left(\frac{\alpha-1}{\alpha}\mathbb{I}-\{B'_0,B'_1\}\right)|\psi'\rangle\right\|\\
    &\leq \frac{2|\alpha-1|}{(3\alpha-1)^{\frac{1}{2}}(\alpha+1)^{\frac{1}{2}}} \delta_4\equiv\delta_5\\
\end{split}
\end{equation}
\begin{equation}\label{anticomR2}
\begin{split}
    \left\|\{X'_A,Z'_A\}|\psi'\rangle\right\|&=\frac{2\alpha}{(\alpha+1)^{\frac{1}{2}}(3\alpha-1)^{\frac{1}{2}}}\left\|\left(\frac{\alpha-1}{\alpha}\mathbb{I}-\{A'_0,A'_1\}\right)|\psi'\rangle\right\|\\
    &\leq  \delta_3+\left\|\{X'_B,Z'_B\}|\psi'\rangle\right\|\\
    &\leq \delta_3+\delta_5\equiv\delta_6
\end{split}
\end{equation}
From \eqref{anticomR2}, we can also get 
\begin{equation}\label{Abound}
    \left\|\left(\{A'_0,A'_1\}-\frac{\alpha-1}{\alpha}\mathbb{I}\right)|\psi'\rangle\right\|\leq \frac{(\alpha+1)^{\frac{1}{2}}(3\alpha-1)^{\frac{1}{2}}}{2\alpha}\delta_6\equiv\delta_7
\end{equation}

Note that $\delta_1\sim\delta_7$ are all of order $O(\sqrt{e})$.

 \item [Step. 3] \textbf{Deriving the bounds for  regularized operators. }
 
Next, we will derive the similar bounds like \eqref{zEquR}, \eqref{xEquR}, \eqref{anticomR} and \eqref{anticomR2} for regularized operators:

\begin{equation}
\begin{split}\label{zEqu}
 \left\|\left(Z_A-Z_B\right)|\psi'\rangle\right\|&=\left\|(Z_A-Z'_B)+(Z'_B-Z_B)|\psi'\rangle\right\|\\
   &\leq \left\|(Z'_A-Z'_B)|\psi'\rangle\right\|+\left \|Z_B(\mathbb{I}-|Z'_B|)\psi'\rangle\right\|\\
   &= \left\|(Z'_A-Z'_B)|\psi'\rangle\right\|+\left \|(\mathbb{I}-|Z'_AZ'_B|)\psi'\rangle\right\|\\
   &\leq\left\|(Z'_A-Z'_B)|\psi'\rangle\right\|+\left \|(\mathbb{I}-Z'_AZ'_B)\psi'\rangle\right\|\\
   &=2\left\|(Z'_A-Z'_B)|\psi'\rangle\right\|=2\delta_1 \equiv\epsilon_1
   \end{split}
\end{equation}
$Z_A=Z'_A$ is an unitary operator. In the forth line, we use the operator inequality $A\leq|A|$.

\begin{equation}
\begin{split}\label{xEqu}
    \left\|\left(X_A-X_B\right)|\psi'\rangle\right\|&\leq\left\|\left(X_A-X'_A\right)|\psi'\rangle\right\|+\left\|\left(X_B-X'_B\right)|\psi'\rangle\right\|+\left\|\left(X'_A-X'_B\right)|\psi'\rangle\right\|\\
    &=\left\|X_A\left(\mathbb{I}-|X'_A|\right)|\psi'\rangle\right\|+\left\|X_B\left(\mathbb{I}-|X'_B|\right)|\psi'\rangle\right\|+\left\|\left(X'_A-X'_B\right)|\psi'\rangle\right\|\\
    &\leq \left\|\left(\mathbb{I}-X'_A+|X'_B|-|X'_B|\right)|\psi'\rangle\right\|+\left\|\left(\mathbb{I}-|X'_B|\right)|\psi'\rangle\right\|+\left\|\left(X'_A-X'_B\right)|\psi'\rangle\right\|\\
    &\leq 2\left\|\left(\mathbb{I}-|X'_B|\right)|\psi'\rangle\right\|+2\left\|\left(X'_A-X'_B\right)|\psi'\rangle\right\|\\
    &\leq 2\left\|\left(\mathbb{I}+|X'_B|\right)\left(\mathbb{I}-|X'_B|\right)|\psi'\rangle\right\|+2\delta_2\\
   &=2\left\|\left(\mathbb{I}-\frac{\alpha^2+(2\alpha-1)^2}{\alpha(3\alpha-1)}\mathbb{I}+\frac{2\alpha-1}{3\alpha-1}\{B'_0,B'_1\}\right)|\psi'\rangle\right\|+2\delta_2\\
   &=\frac{2|2\alpha-1|}{3\alpha-2}\left\|\left(\{B'_0,B'_1\}-\frac{\alpha-1}{\alpha}\mathbb{I}\right)|\psi'\rangle\right\|+2\delta_2\\
   &\leq \frac{2|2\alpha-1|}{3\alpha-2}\delta_4+2\delta_2\equiv\epsilon_2
\end{split}
\end{equation}
 In the fifth line, we use the operator inequality $\mathbb{I}+|A|\geq\mathbb{I}$.
 \begin{equation}
\begin{split}\label{Aanti}
    \left\|\{X_A,Z_A\}|\psi'\rangle\right\|&\leq\left\|X_A(Z_A-Z'_A)|\psi'\rangle\right\|+\left\|Z_A(X_A-X'_A)|\psi'\rangle\right\|+\left\|(Z_A-Z'_A)X'_A|\psi'\rangle\right\|\\
    &+\left\|(X_A-X'_A)Z'_A|\psi'\rangle\right\|+ \left\|\{X'_A,Z'_A\}|\psi'\rangle\right\|\\
    &=\left\|(\mathbb{I}-|X'_A|)|\psi'\rangle\right\|+\left\|(\mathbb{I}-|X'_A|)Z'_A|\psi'\rangle\right\|+ \left\|\{X'_A,Z'_A\}|\psi'\rangle\right\|\\
    &\leq \frac{|2\alpha-1|}{3\alpha-2}\delta_4+\delta_2+\delta_6+\left\|(\mathbb{I}-{X'_A}^2)Z'_A|\psi'\rangle\right\|\\
    &= \frac{|2\alpha-1|}{3\alpha-2}\delta_4+\delta_2+\delta_6\\
    &\quad+\frac{2|\alpha-1|}{\alpha(\alpha+1)(3\alpha-1)}\left\|\left(\{A'_0,A'_1\}-\frac{\alpha-1}{\alpha}\mathbb{I}\right)Z'_A|\psi'\rangle\right\|\\
    &\leq \frac{|2\alpha-1|}{3\alpha-2}\delta_4+\delta_2+\delta_6+\frac{2|\alpha-1|}{\alpha(\alpha+1)(3\alpha-1)}\delta_7 \equiv\epsilon_3
\end{split}
\end{equation}
In the third line, we use the intermediate result of \eqref{xEqu} and play the same trick as line 5 of \eqref{xEqu}. In the last line, we use the bound \eqref{Abound}.
\begin{equation}
\begin{split}\label{Banti}
    \left\|\{X_B,Z_B\}|\psi'\rangle\right\|&\leq\left\|X_B(Z_B-Z_A)|\psi'\rangle\right\|+\left\|Z_B(X_B-X_A)|\psi'\rangle\right\|+\left\|X_A(Z_B-Z_A)|\psi'\rangle\right\|\\
    &\quad+\left\|Z_A(X_B-X_A)|\psi'\rangle\right\|
    + \left\|\{X_A,Z_A\}|\psi'\rangle\right\|    \\
    &\leq2\epsilon_1+2\epsilon_2+\epsilon_3\equiv\epsilon_4
\end{split}
\end{equation}

Similarly, $\epsilon_1\sim\epsilon_4$ are all of order $O(\sqrt{e})$.
\item [Step. 4] \textbf{Using the derived bounds to prove the robust self-testing statements. }

With these conditions \eqref{zEqu},  \eqref{xEqu}, \eqref{Aanti} and \eqref{Banti}, it's easy to see the following statement holds:
\begin{equation}\label{robust}
    \left\|\Phi\left(A'_x\otimes B'_y|\psi'\rangle\right)-\left(A_x\otimes B_y|\phi^+\rangle\right)\otimes|junk\rangle\right\| \leq  O(\sqrt{e})
\end{equation}
where $x,y\in\{-1,0,1\}$, the subscript -1 refers to $A_{-1}=B_{-1}=\mathbb{I}$ and the isometry is shown as Fig.~\ref{swap}. The output of the isometry is:
\begin{equation}
\begin{split}\label{isometry}
    \Phi\left(A'_x\otimes B'_y|\psi'\rangle\right)=& \frac{1}{4}\left[|00\rangle \otimes(\mathbb{I}+Z_A)(\mathbb{I}+Z_B)A'_x \otimes B'_y|\psi'\rangle\right.\\
    &+|01\rangle\otimes X_B(\mathbb{I}+Z_A)(\mathbb{I}-Z_B)A'_x \otimes B'_y|\psi'\rangle\\
    &+|10\rangle\otimes X_A(\mathbb{I}-Z_A)(\mathbb{I}+Z_B)A'_x \otimes B'_y|\psi'\rangle\\
    &\left.+|11\rangle\otimes X_AX_B(\mathbb{I}-Z_A)(\mathbb{I}-Z_B)A'_x\otimes B'_y|\psi'\rangle\right]\\
\end{split}
\end{equation}
\begin{enumerate}
    \item [1.] \textbf{Robust self-testing the state $|\psi'\rangle$.}
    
When $x=y=-1$, the second and third terms of \eqref{isometry} are bounded by:
\begin{equation}
\begin{split}
     \left\|(\mathbb{I}+Z_A)(\mathbb{I}-Z_B)|\psi'\rangle\right\|&=\left\|(\mathbb{I}-Z_A)(\mathbb{I}+Z_B)|\psi'\rangle\right\|\\
     &\leq \left\|\left(Z_A-Z_B\right)|\psi'\rangle\right\|+\left\|\left(\mathbb{I}-Z_AZ_B\right)|\psi'\rangle\right\|
    \\
    &\leq 2\left\|\left(Z_A-Z_B\right)|\psi'\rangle\right\|=2\epsilon_1   
\end{split}
\end{equation}
The difference of the first and fourth terms of \eqref{isometry} is:
\begin{equation}
\begin{split}
        &\left\|\left[X_AX_B(\mathbb{I}-Z_A)(\mathbb{I}-Z_B)-(\mathbb{I}+Z_A)(\mathbb{I}+Z_B)\right]|\psi'\rangle\right\|\\
        &\leq \left\|\left[(\mathbb{I}+Z_A)(\mathbb{I}+Z_B)(X_AX_B-\mathbb{I})\right]|\psi'\rangle\right\|\\
        &\quad+2\left\|\{X_B,Z_B\}|\psi'\rangle\right\|+2\left\|\{X_A,Z_A\}|\psi'\rangle\right\|\\
        &\leq 4\left\|\left(X_A-X_B\right)|\psi'\rangle\right\|+2(\epsilon_3+\epsilon_4)\\&=4\epsilon_2+2\epsilon_3+2\epsilon_4
\end{split}
\end{equation}
Thus $\left\|\Phi\left(|\psi'\rangle\right)-|\phi^+\rangle\otimes\frac{\sqrt{2}}{4}(\mathbb{I}+Z_A)(\mathbb{I}+Z_B)|\psi'\rangle\right\| \leq4\epsilon_1+4\epsilon_2+2\epsilon_3+2\epsilon_4=O(\sqrt{e})$.

 \item [2.] \textbf{Robust self-testing the local measurements $A'_x$, $x\in\{0,1\}$ on Alice's side.}
 
Since $A'_x$, $x\in\{0,1\}$ is the linear combination of $Z'_A$ and $X'_A$, here we take the local operator to be either $Z'_A$ or $X'_A$, and the corresponding reference operators is $\sigma_z^A$ or $\sigma_x^A$ (the Pauli operators).

When $A'_x=Z'_A,y=-1$, the second and third terms of \eqref{isometry} are bounded by:

\begin{equation}
\begin{split}
    \left\|X_B(\mathbb{I}+Z_A)(\mathbb{I}-Z_B)Z'_A|\psi'\rangle\right\|&=
     \left\|X_A(\mathbb{I}-Z_A)(\mathbb{I}+Z_B)Z'_A|\psi'\rangle\right\|\\
     &=\left\|(\mathbb{I}+Z_A)(\mathbb{I}-Z_B)|\psi'\rangle\right\|\leq 2\epsilon_1.
\end{split}
\end{equation}

The difference of the  fourth  term and the negative of the first term in \eqref{isometry} is:
\begin{equation}
\begin{split}
   &\left\|\left[X_AX_B(\mathbb{I}-Z_A)(\mathbb{I}-Z_B)Z'_A-\left(-(\mathbb{I}+Z_A)(\mathbb{I}+Z_B)Z'_A\right)\right]|\psi'\rangle\right\|\\&=\left\|\left[(\mathbb{I}+Z_A)(\mathbb{I}+Z_B)-X_AX_B(\mathbb{I}-Z_A)(\mathbb{I}-Z_B)\right]|\psi'\rangle\right\|\\
   &\leq 4\epsilon_2+2\epsilon_3+2\epsilon_4.
\end{split}
\end{equation}

Thus $\left\|\Phi\left(Z'_A|\psi'\rangle\right)-\sigma_z^A|\phi^+\rangle\otimes\frac{\sqrt{2}}{4} (\mathbb{I}+Z_A)(\mathbb{I}+Z_B)Z'_A|\psi'\rangle\right\| \leq
4\epsilon_1+4\epsilon_2+2\epsilon_3+2\epsilon_4=O(\sqrt{e})$.

For $A'_x=X'_A,y=-1$, we will use the fact that $\Phi$ is linear and preservers the norms, $\left\|\Phi\left(X'_A|\psi'\rangle\right)-\Phi\left(X_A|\psi'\rangle\right)\right\|=\left\|\left(X'_A-X_A\right)|\psi'\rangle\right\|=\frac{\epsilon_2}{2}$. This allows us to bound the \eqref{robust} with $A'_x=X'_A,y=-1$ by  $A'_x=X_A,y=-1$.


When $A'_x=X_A,y=-1$, the difference of the second and third terms of \eqref{isometry} is bounded by:
\begin{equation}
\begin{split}
    &\left\|\left[ X_B(\mathbb{I}+Z_A)(\mathbb{I}-Z_B)X_A
    -X_A(\mathbb{I}-Z_A)(\mathbb{I}+Z_B)X_A\right]|\psi'\rangle\right\|\\&\leq  
    \left\| (\mathbb{I}+Z_A)(\mathbb{I}+Z_B)(X_AX_B-\mathbb{I})|
    \psi'\rangle\right\|\\
    &\quad+2\left\|\{X_B,Z_B\}X_A|\psi'\rangle\right\|+2\left\|\{X_A,Z_A\}X_A|\psi'\rangle\right\|\\
    &\leq4\left\|\left(X_A-X_B\right)|\psi'\rangle\right\|+2\left\|\{X_B,Z_B\}|\psi'\rangle\right\|+2\left\|\{X_A,Z_A\}|\psi'\rangle\right\|\\
    &\leq 4\epsilon_2+2\epsilon_3+2\epsilon_4.
\end{split}
\end{equation}
And the first term and the fourth term of \eqref{isometry} are bounded by:
{\small
\begin{equation}
\begin{split}
    \left\|(\mathbb{I}+Z_A)(\mathbb{I}+Z_B)X_A|\psi'\rangle\right\|
    &\leq 
    \left\|\{X_A,Z_A\}(\mathbb{I}+Z_B)|\psi'\rangle\right\|+\left\|X_A(\mathbb{I}-Z_A)(\mathbb{I}+Z_B)|\psi'\rangle\right\|\\
    &\leq 2\left\|\{X_A,Z_A\}|\psi'\rangle\right\|+\left\|(\mathbb{I}-Z_A)(\mathbb{I}+Z_B)|\psi'\rangle\right\|\\
    &\leq 2\epsilon_1 +2\epsilon_3,\\
   \left\|X_AX_B(\mathbb{I}-Z_A)(\mathbb{I}-Z_B)X_A|\psi'\rangle\right\|&\leq
  \left\|X_B(\mathbb{I}+Z_A)(\mathbb{I}-Z_B)|\psi'\rangle\right\|+\left\|X_AX_B\{X_A,Z_A\}(\mathbb{I}-Z_B)|\psi'\rangle\right\|\\
  &\leq \left\|(\mathbb{I}+Z_A)(\mathbb{I}-Z_B)|\psi'\rangle\right\|+2\left\|\{X_A,Z_A\}|\psi'\rangle\right\|\\
  &\leq2\epsilon_1 +2\epsilon_3.
\end{split}
\end{equation}}

Thus $\left\|\Phi\left(X_A|\psi'\rangle\right)-\sigma_x^A|\phi^+\rangle\otimes\frac{\sqrt{2}}{4} X_B(\mathbb{I}+Z_A)(\mathbb{I}-Z_B)X_A|\psi'\rangle\right\| \leq4\epsilon_1 +4\epsilon_2+6\epsilon_3+2\epsilon_4.
$. Then:
\begin{equation}
\begin{split}
    &\left\|\Phi\left(X'_A|\psi'\rangle\right)-\sigma_x^A|\phi^+\rangle\otimes\frac{\sqrt{2}}{4} X_B(\mathbb{I}+Z_A)(\mathbb{I}-Z_B)X_A|\psi'\rangle\right\|\\&\leq4\epsilon_1 +4\epsilon_2+6\epsilon_3+2\epsilon_4+\left\|\Phi\left(X'_A|\psi'\rangle\right)-\Phi\left(X_A|\psi'\rangle\right)\right\|\\
    &\leq 4\epsilon_1 +\frac{9}{2}\epsilon_2+6\epsilon_3+2\epsilon_4=O(\sqrt{e})
\end{split}
\end{equation}

\item [3.] \textbf{Robust self-testing the local measurements $B'_y$, $y\in\{0,1\}$ on Bob's side .}

The proof is similar as the above case on Alice's side.
\end{enumerate}

\end{enumerate}
\end{proof}

\section{The guessing probability of a classical adversary allowed to prepare arbitrary no-signaling boxes for Alice and Bob}\label{bound_prob}
Suppose Alice and Bob share an arbitrary bipartite no-signaling box $P_{A,B|X,Y}$ that achieves a value $I_{\alpha}^{Hardy}$ for the Bell expression \eqref{eq:I-Hardy} derived from the tilted Hardy test. The following proposition derives a bound on the probability $P_{A,B|X,Y}(a,b|A_x,B_y)$ for any pair of outcomes $a,b$ and any pair of inputs $x,y$. The corresponding bound for the MDL quantity $I_{\alpha}^{MDL}$ is then readily derived by the relationship between these two quantities that $I_{\alpha}^{Hardy} \geq \frac{I_{\alpha}^{MDL}}{l h}$ (see Eq.~\eqref{MDL} ) for $l,h$ being the lower and upper bounds on the probability of choosing measurement inputs $x,y$ respectively. Note that when measurement inputs are chosen using an $\epsilon$-SV source, we have $l = \left(\frac{1}{2} - \epsilon \right)^2$ and $h = \left( \frac{1}{2} + \epsilon \right)^2$. 

\begin{proposition*}
Consider any bipartite no-signaling box $P_{A,B|X,Y}$ that achieves the value $I_{\alpha}^{Hardy}$ for the expression in \eqref{eq:I-Hardy}. For any pair of measurement settings $x, y$ and outcomes $a, b$, the following upper bound holds:
\begin{equation}
    P_{A,B|X,Y}(a,b|A_x, B_y) \leq 1-\frac{1}{\alpha}I_{\alpha}^{Hardy} \qquad \text{for } \alpha \in (\frac{1}{3},1]
\end{equation}
and
\begin{equation}
    P_{A,B|X,Y}(a,b|A_x, B_y) \leq 1-I_{\alpha}^{Hardy} \qquad \text{for } \alpha \in (1,+\infty).
\end{equation}

\end{proposition*}

\begin{proof}
In the bipartite Bell scenario with each party choosing two binary measurements Fig.~\ref{propagation} (a) illustrates the following relations:
\begin{equation}
    \begin{split}
        P_{A,B|X,Y}(0,0|A_0,B_0)+x&=c+\delta_1,\\
        d&\geq c-\delta_2 ,\\
        e'&\geq d-\delta_3,\\
        P_{A,B|X,Y}(1,1|A_0,B_0)&\geq e'-x.
    \end{split}
\end{equation}

We derive 
\begin{equation}\label{relation_block1}
  P_{A,B|X,Y}(1,1|A_0,B_0)\geq P_{A,B|X,Y}(0,0|A_0,B_0)-\delta_1-\delta_2-\delta_3.
\end{equation}

In the following, we will discuss the two cases $\alpha\in(\frac{1}{3},1]$ and $\alpha\in(1,+\infty)$ respectively.
\begin{description}
    \item[Case 1.] $\alpha \in (\frac{1}{3},1]$. 
    In this case, we observe  that $\max(0,\frac{1-\alpha}{2}) = \frac{1-\alpha}{2}$ and derive the bounds
    \begin{align}
   & \frac{1-\alpha}{2}+I_{\alpha}^{Hardy} \leq P_{A,B|X,Y}(0,0|A_0,B_0) \leq \frac{1+\alpha}{2} - I_{\alpha}^{Hardy}, \nonumber \\
   & \frac{1-\alpha}{2}+I_{\alpha}^{Hardy} \leq P_{A,B|X,Y}(1,1|A_0,B_0) \leq \frac{1+\alpha}{2} - I_{\alpha}^{Hardy},\nonumber \\
   & \frac{1}{\alpha}I_{\alpha}^{Hardy} \leq P_{A,B|X,Y}(a,b|A_x, B_y) \leq 1-\frac{1}{\alpha}I_{\alpha}^{Hardy}, \quad \forall a\oplus b=x\cdot y,((x,y)\neq (0,0)), \nonumber \\
   &  P_{A,B|X,Y}(a,b|A_x, B_y) \leq 1-\frac{2}{\alpha}I_{\alpha}^{Hardy},\quad \forall a\oplus b\neq x\cdot y.
    \end{align}    
    Firstly, we show that the following bounds hold:
        
        From the definition of $I_{\alpha}^{Hardy}$ and the normalization constraint $ P_{A,B|X,Y}(0,0|A_0,B_0)+P_{A,B|X,Y}(1,1|A_0,B_0)\leq 1$, we can get:
        \begin{equation}
        \begin{split}
          P_{A,B|X,Y}(0,0|A_0,B_0)&\geq \frac{1-\alpha}{2}+I_{\alpha}^{Hardy}\\ P_{A,B|X,Y}(1,1|A_0,B_0)&\leq \frac{1+\alpha}{2}-I_{\alpha}^{Hardy}
        \end{split}
        \end{equation}
        
        In order to calculate the bounds of other probabilities, from Fig.~\ref{propagation} (b), we get the following inequalities:
\begin{equation}
\begin{split}
    P_{A,B|X,Y}(0,0|A_0,B_0)-\delta_1\leq c\\
    P_{A,B|X,Y}(0,0|A_0,B_0)-\delta_3 \leq e\\
\end{split}
\end{equation}
Since $c-\delta_2\leq d, e-\delta_2\leq d'$, we  get:
\begin{equation}
\begin{split}
        P_{A,B|X,Y}(0,0|A_0,B_0)-\delta_1-\delta_2 \leq d\\
        P_{A,B|X,Y}(0,0|A_0,B_0)-\delta_2-\delta_3 \leq d'
\end{split}
\end{equation}
Since $d-\delta_3\leq e',d'-\delta_1\leq c'$, we get:
\begin{equation}
\begin{split}
        P_{A,B|X,Y}(0,0|A_0,B_0)-\delta_1-\delta_2-\delta_3\leq e'\\
        P_{A,B|X,Y}(0,0|A_0,B_0)-\delta_1-\delta_2-\delta_3\leq c'
\end{split}
\end{equation}
Thus, combining Eq.~\eqref{relation_block1} and the definition of $I_{\alpha}^{Hardy}$, these probabilities $c,c',d,d',e,e'$ are larger than or equal to $\frac{1}{\alpha}I_{\alpha}^{Hardy}$. Using the normalization constraints for different measurement settings, one can get corresponding upper bounds for these probabilities. For example, from $c+c'+\delta_1+s=1$, one can get the upper bounds of $c,c'$:
\begin{equation}
\begin{split}
      c&\leq 1-c'-\delta_1\leq 1-P_{A,B|X,Y}(0,0|A_0,B_0)+\delta_2+\delta_3\\
      c'&\leq 1-c-\delta_1\leq 1-P_{A,B|X,Y}(0,0|A_0,B_0)
\end{split}
\end{equation}
Thus, the entries $c,c',d,d',e,e'$ are smaller than or equal to $1-\frac{1}{\alpha}I_{\alpha}^{Hardy}$. For entry $\delta_i$, it is simply bounded by $\delta_i\leq \delta_1+\delta_2+\delta_3$. From Fig.~\ref{propagation} (c), one can obtain the following inequalities:
\begin{equation}
\begin{split}
        P_{A,B|X,Y}(0,0|A_0,B_0)-\delta_1-\delta_2\leq d\\
        P_{A,B|X,Y}(0,0|A_0,B_0)-\delta_2-\delta_3\leq d'
\end{split}
\end{equation}
Together with $d+d'+\delta_2\leq 1$, we get:
\begin{equation}
    P_{A,B|X,Y}(0,0|A_0,B_0)\leq \frac{1}{2}(1+\delta_1+\delta_2+\delta_3).
\end{equation}
Thus, combining Eq.~\eqref{relation_block1} and the definition of $I_{\alpha}^{Hardy}$, we get:
\begin{equation}
    \delta_1+\delta_2+\delta_3\leq \alpha-2I_{\alpha}^{Hardy}
\end{equation}
 and 
 \begin{equation}
    P_{A,B|X,Y}(0,0|A_0,B_0)\leq \frac{1+\alpha}{2}-I_{\alpha}^{Hardy}.
 \end{equation}
Together with the normalization constraint, we get $P_{A,B|X,Y}(1,1|A_0,B_0)\geq \frac{1-\alpha}{2}+I_{\alpha}^{Hardy}$.
For the remaining entries $x,x',s,s',t$, their upper bounds can directly found by using no-signaling or normalization conditions as shown in Fig.~\ref{propagation} (d).
In general, we obtain the upper bound
    \begin{equation}
        P_{A,B|X,Y}(a,b|A_x,B_y)\leq 1-\frac{1}{\alpha}I_{\alpha}^{Hardy},
    \end{equation}
    for $w \in (\frac{1}{3}, 1 ]$.
    
  \item[Case 2.] $\alpha \in (1, +\infty)$. 
  In this case, we observe that $\max(0,\frac{1-\alpha}{2}) = 0$ and derive the bounds
    \begin{align}
    I_{\alpha}^{Hardy} \leq P_{A,B|X,Y}(a,b|A_x, B_y) &\leq 1-I_{\alpha}^{Hardy} \qquad \forall a\oplus b=x\cdot y  \nonumber \\
     P_{A,B|X,Y}(a,b|A_x, B_y) &\leq 1-2I_{\alpha}^{Hardy}\qquad \forall a\oplus b\neq x\cdot y
    \end{align}

    The proof in this case is similar to the proof of  Case 1, we omit the details here.
     For any given $\alpha \in (1,+\infty)$, we obtain the upper bound
    \begin{equation}
        P_{A,B|X,Y}(a,b|A_x, B_y) \leq 1-I_{\alpha}^{Hardy}
    \end{equation}
\end{description}

\begin{figure}[H]
  \centering
  \includegraphics[width=0.6\textwidth]{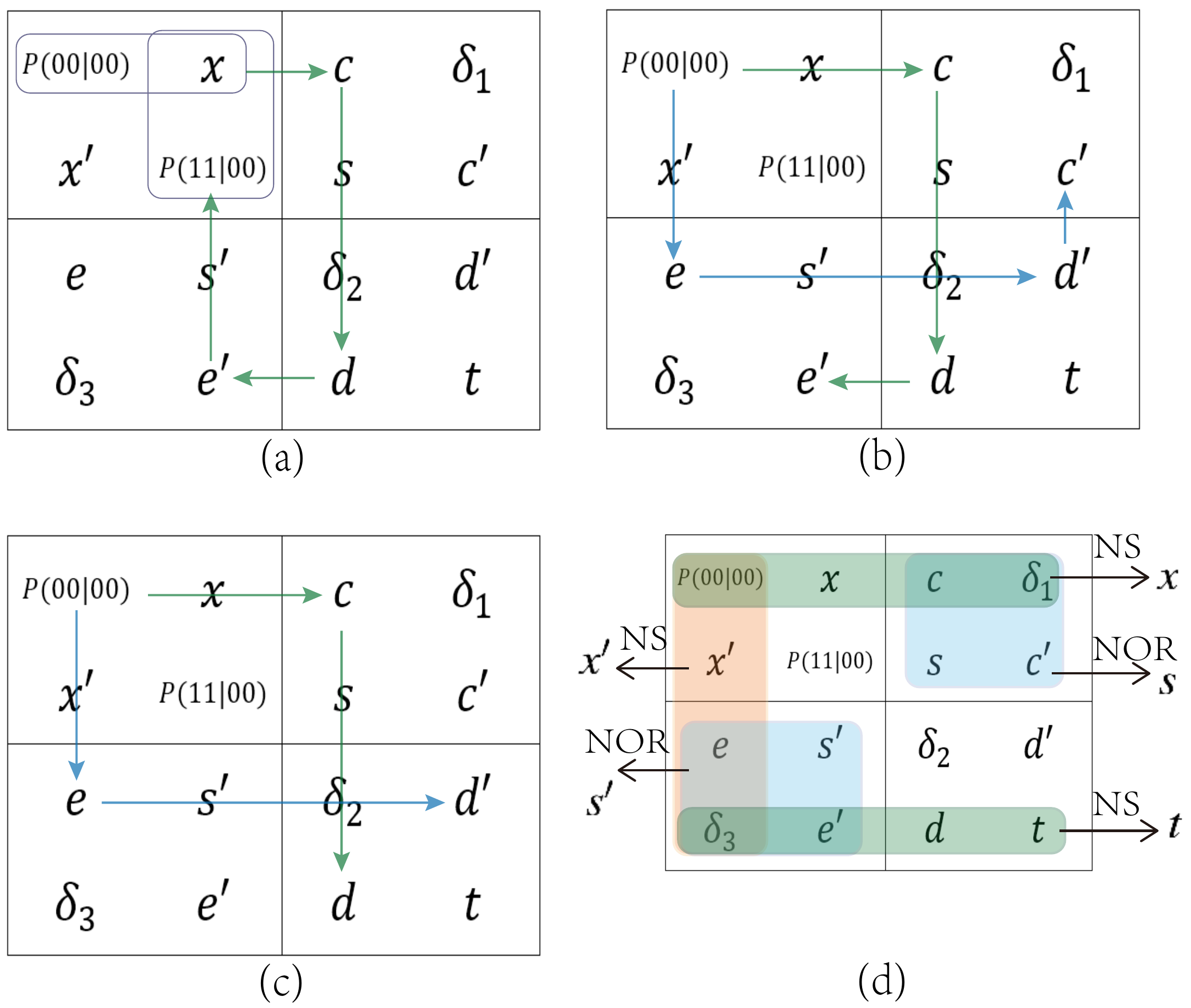}\\
  \caption{(a), (b) and (c): Chains of inequalities under the no-signaling conditions. (d): No-signaling and normalization constraints used for calculating the upper bounds of $x,x',s,s',t$. (NS: no-signaling constraint; NOR: normalization constraint.)}\label{propagation}
\end{figure}
\end{proof}

\end{appendix}

\end{document}